\documentclass[prl,twocolumn,aps,psfig,showpacs,nofootinbib,nobibnotes,superscriptaddress,preprintnumbers,times]{revtex4}
\usepackage{graphicx,bm,color}
\graphicspath{{./fig/}{./png/}}

\newcommand{\EQ}{\begin{equation}}
\newcommand{\EN}{\end{equation}}
\newcommand{\const}{{\rm const}{}}

\newcommand{\rr}{\bm{r}}
\newcommand{\uu}{\bm{u}}

\newcommand{\BB}{\bm{B}}
\newcommand{\AAA}{\bm{A}}

\newcommand{\xx}{\bm{x}}
\newcommand{\nab}{\mbox{\boldmath $\nabla$} {}}

\newcommand{\SSSS}{\mbox{\boldmath ${\sf S}$} {}}

\def\urms{u_{\rm rms}}
\def\orms{\omega_{\rm rms}}
\def\cs{c_{\rm s}}
\def\csz{c_{\rm s0}}

\def\dd{{\rm d}}
\def\sgn{{\rm sgn}}
\newcommand{\bra}[1]{\langle #1\rangle}
\newcommand{\Fig}[1]{Fig.~\ref{#1}}

\newcommand{\Figp}[2]{Fig.~\ref{#1}({#2})}
\newcommand{\Figsp}[3]{Figs.~\ref{#1}({#2}) and ({#3})}
\newcommand{\Figssp}[3]{Figs.~\ref{#1}({#2})--({#3})}

\newcommand{\Tab}[1]{Table~\ref{#1}}

\def\EK{E_{\rm K}}
\def\EM{E_{\rm M}}

\def\Pm{\mbox{\rm Pr}_{\rm M}}

\def\Rey{\mbox{\rm Re}}

\def\pM{p_{\rm M}}
\def\qM{q_{\rm M}}

\newcommand{\arXiv}[2]{, arXiv:#2 (#1).}
\newcommand{\yjcap}[3]{, J.\ Cosmol.\ Astropart.\ Phys. {\bf #2} (#1) #3.}
\newcommand{\yprd}[3]{, Phys.\ Rev.\ D {\bf #2}, #3 (#1).}
\newcommand{\ypre}[3]{, Phys.\ Rev.\ E {\bf #2}, #3 (#1).}
\newcommand{\yprl}[3]{, Phys.\ Rev.\ Lett.\ {\bf #2}, #3 (#1).}
\newcommand{\yjfm}[3]{, J.\ Fluid Mech.\ {\bf #2}, #3 (#1).}

\newcommand{\eq}[1]{(\ref{#1})}

\newcommand{\Eq}[1]{Eq.~(\ref{#1})}

\newcommand{\yjour}[4]{, #2 {\bf #3}, #4 (#1).}

\newcommand{\ypf}[3]{, Phys. Fluids {\bf #2}, #3 (#1).}

\newcommand{\yapj}[3]{, Astrophys. J. {\bf #2}, #3 (#1).}

\newcommand{\ymn}[3]{, Mon.\ Not.\ R.\ Astron.\ Soc.\ {\bf #2}, #3 (#1).}
\newcommand{\yptrsa}[3]{, Phil. Trans. Roy. Soc. Lond. A, {\bf #2}, #3 (#1).}
\newcommand{\yproc}[4]{, (ed. #4), pp. #2. #3 (#1).}

\begin{document}

\title{Classes of hydrodynamic and magnetohydrodynamic turbulent decay}

\date{Received 5 July 2016; published 3 February 2017,~ $ $Revision: 1.66 $ $}
\preprint{NORDITA-2016-82 --- Phys. Rev. Lett. 118, 055102 (2017)}

\author{Axel Brandenburg}
\email{brandenb@nordita.org}
\affiliation{Laboratory for Atmospheric and Space Physics, University of Colorado, Boulder, CO 80303, USA}
\affiliation{JILA and Department of Astrophysical and Planetary Sciences, University of Colorado, Boulder, CO 80303, USA}
\affiliation{Nordita, KTH Royal Institute of Technology and Stockholm University,
Roslagstullsbacken 23, 10691 Stockholm, Sweden}
\affiliation{Department of Astronomy, AlbaNova University Center,
Stockholm University, 10691 Stockholm, Sweden}

\author{Tina Kahniashvili}
\email{tinatin@andrew.cmu.edu}
\affiliation{The McWilliams Center for
Cosmology and Department of Physics, Carnegie Mellon University,
5000 Forbes Ave, Pittsburgh, PA 15213, USA}
\affiliation{Department of Physics, Laurentian University, Ramsey
Lake Road, Sudbury, ON P3E 2C, Canada} \affiliation{Abastumani Astrophysical Observatory, Ilia State
University, 3-5 Cholokashvili Ave, Tbilisi, GE-0194, Georgia}

\begin{abstract}
We perform numerical simulations of decaying hydrodynamic and
magnetohydrodynamic turbulence.
We classify our time-dependent solutions by their evolutionary tracks
in parametric plots between instantaneous scaling exponents.
We find distinct classes of solutions evolving along specific
trajectories toward points on a line of self-similar solutions.
These trajectories are determined by the underlying physics
governing individual cases, while the infrared slope of the
initial conditions plays only a limited role.
In the helical case, even for a scale-invariant initial spectrum
(inversely proportional to wavenumber $k$), the solution evolves
along the same trajectory as for a Batchelor spectrum
(proportional to $k^4$).
\end{abstract}

\maketitle

The study of {\em decaying} turbulence is as old as that of turbulence itself.
Being independent of an ill-defined forcing mechanism, decaying turbulence
has a better chance in displaying generic properties of turbulence.
Such properties are usually reflected in the existence of conserved
quantities such as the Loitsiansky integral \cite{BP56} and the
magnetic helicity \cite{PFL76,BM99}.
Important applications of decaying turbulence include grid turbulence
\cite{SSD99}, turbulent wakes \cite{Cas70}, atmospheric turbulence
\cite{Lil83}, as well as interstellar turbulence \cite{MLKB98},
galaxy clusters \cite{SSH06}, and the early Universe \cite{CHB01,BJ04}.
In the latter case, cosmological magnetic fields generated in the early
Universe provide the initial source of turbulence, which leads to a growth
of the correlation length by an inverse cascade mechanism \cite{BEO96},
in addition to the general cosmological expansion of the Universe.
In the last two decades, this topic has gained significant attention
\cite{Cam15}.
The time span since the initial magnetic field generation is enormous,
but it is still uncertain whether it is long enough to produce fields
at sufficiently large length scales to explain the possibility of
contemporary magnetic fields in the space between clusters of galaxies
\cite{WB16}.

In this Letter, we use direct numerical simulations (DNS) of both
hydrodynamic (HD) and magnetohydrodynamic (MHD) decaying turbulence to
classify different types by their decay behavior.
The decay is characterized by the temporal change of the kinetic
energy spectrum, $\EK(k,t)$, and, in MHD, also by the magnetic energy
spectrum, $\EM(k,t)$.
Here, $k$ is the wavenumber and $t$ is time.
In addition to the decay laws of the energies
${\cal E}_i(t)=\int E_i(k,t)\,dk$, with $i={\rm K}$ or ${\rm M}$ for
kinetic and magnetic energies, there are the kinetic and magnetic
integral scales,
\EQ
\xi_i(t)=\left.\int_0^\infty k^{-1} E_i(k,t)\,dk\right/
\int_0^\infty E_i(k,t)\,dk.
\label{xirelation2}
\EN
We quantify the decay by the instantaneous scaling exponents
$p(t)\equiv d\ln{\cal E}/d\ln t$ and $q(t)\equiv d\ln\xi/d\ln t$.
Thus, we study the decay behaviors by plotting $p(t)$ vs.\ $q(t)$
in a parametric representation.
The $pq$ diagram turns out to be a powerful diagnostic tool.

Earlier work \cite{SSH06,Geo92,Ole97} has suggested that the decay
behavior, and thus the positions of solutions in the $pq$ diagram,
depend on the exponent $\alpha$ for initial conditions of the form
$E\sim k^\alpha e^{-k/k_0}$, where $k_0$ is a cutoff wavenumber.
Motivated by earlier findings \cite{PFL76,BEO96} of an inverse cascade
in decaying MHD turbulence, Olesen considered the time-dependent energy
spectra $E(k,t)$ to be of the form \cite{Ole97}
\EQ
E(k,t)\propto k^\alpha \psi\left(k\xi(t)\right),
\label{scalinglaw}
\EN
where $\xi(t)\propto t^q$, with $q$ being an as yet undetermined
scaling exponent, and $\psi$ is a function that depends on the dissipative
and turbulent processes that lead to a departure from a
powerlaw at large $k$.
Moreover, the slope $\psi'\equiv d\psi/d\kappa$ with $\kappa=k\xi$ must
vanish for $\kappa\to0$.
This turns out to be a critical restriction.

Olesen then makes use of the fact that the HD and MHD
equations are invariant under rescaling, $x\to\tilde{x}\ell$
and $t\to\tilde{t}\ell^{1/q}$, which implies corresponding
rescalings for velocity $u\to\tilde{t}\ell^{1-1/q}$ and viscosity
$\nu\to\tilde{\nu}\ell^{2-1/q}$.
Furthermore, using the fact that the dimensions of $E(k,t)$ are given by
$[E]=[x]^3[t]^{-2}$, and requiring $\psi$ to be invariant under rescaling
$E\to\tilde{E}\ell^{3-2/q}\propto\tilde{k}^\alpha\ell^{-\alpha}\psi$,
he finds from \Eq{scalinglaw} that $\alpha=-3+2/q$.
He argues that for a given subinertial range spectral exponent
$\alpha$, the exponent $q$ is given by \cite{Cam15,Ole97,KP04,Cam04}
\EQ
q=2/(3+\alpha)
\label{qrelation}
\EN
for both HD and MHD and independent of the presence or absence of helicity.
A remarkable prediction of Olesen's original work concerns the existence of inverse
transfer even in the absence of magnetic helicity, provided $\alpha>-3$.
In subsequent work he stresses that for constant $\nu$ (and $\eta$), only
the case $\alpha=1$ can be realized.
For nonhelical MHD, this is indeed compatible with simulations
\cite{BKT15,Zra14,Ole15a}, but not for HD \cite{Kang} nor for helical MHD
\cite{BM99,TKBK12}.

\begin{figure*}[t!]
\includegraphics[width=.32\textwidth]{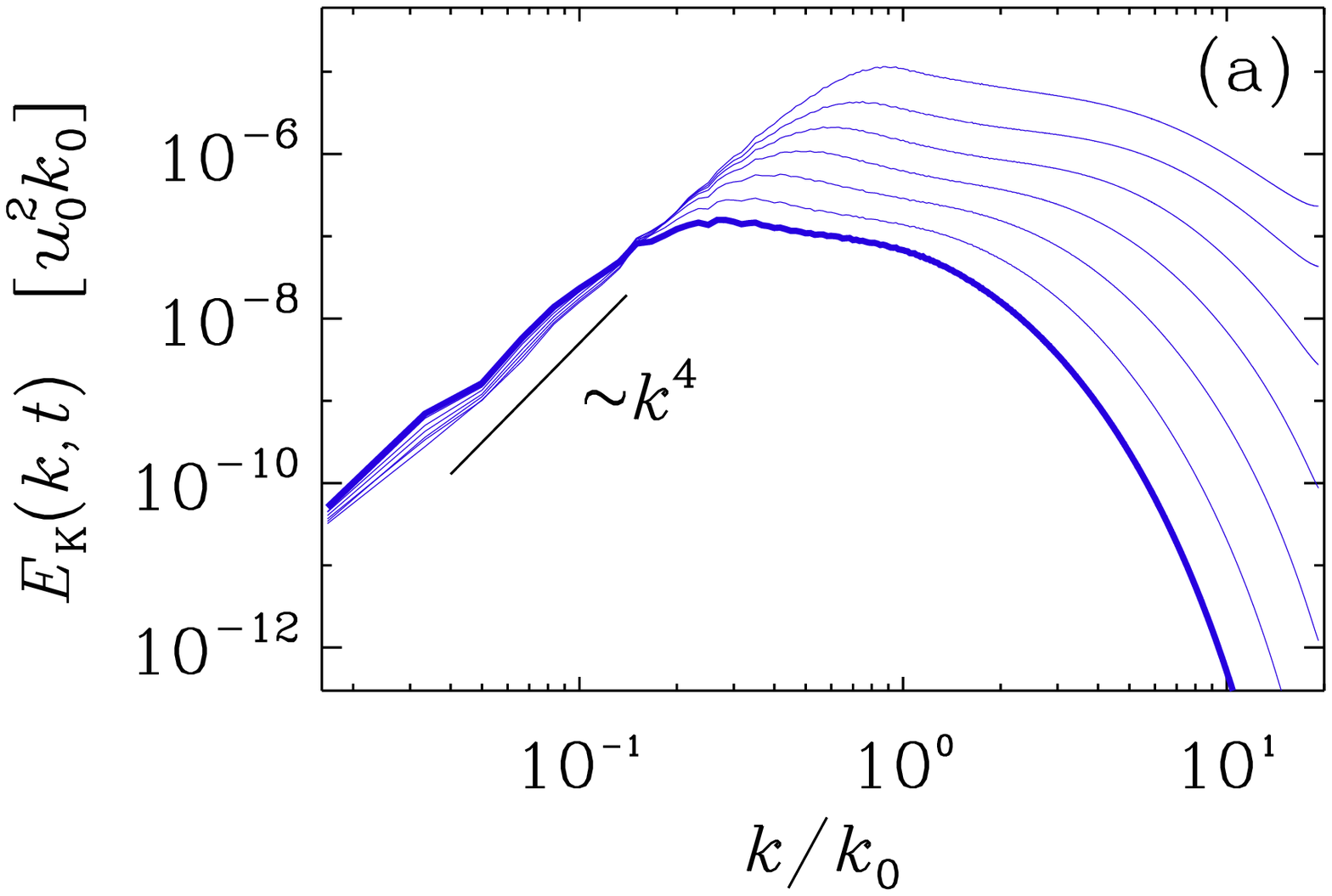}
\includegraphics[width=.32\textwidth]{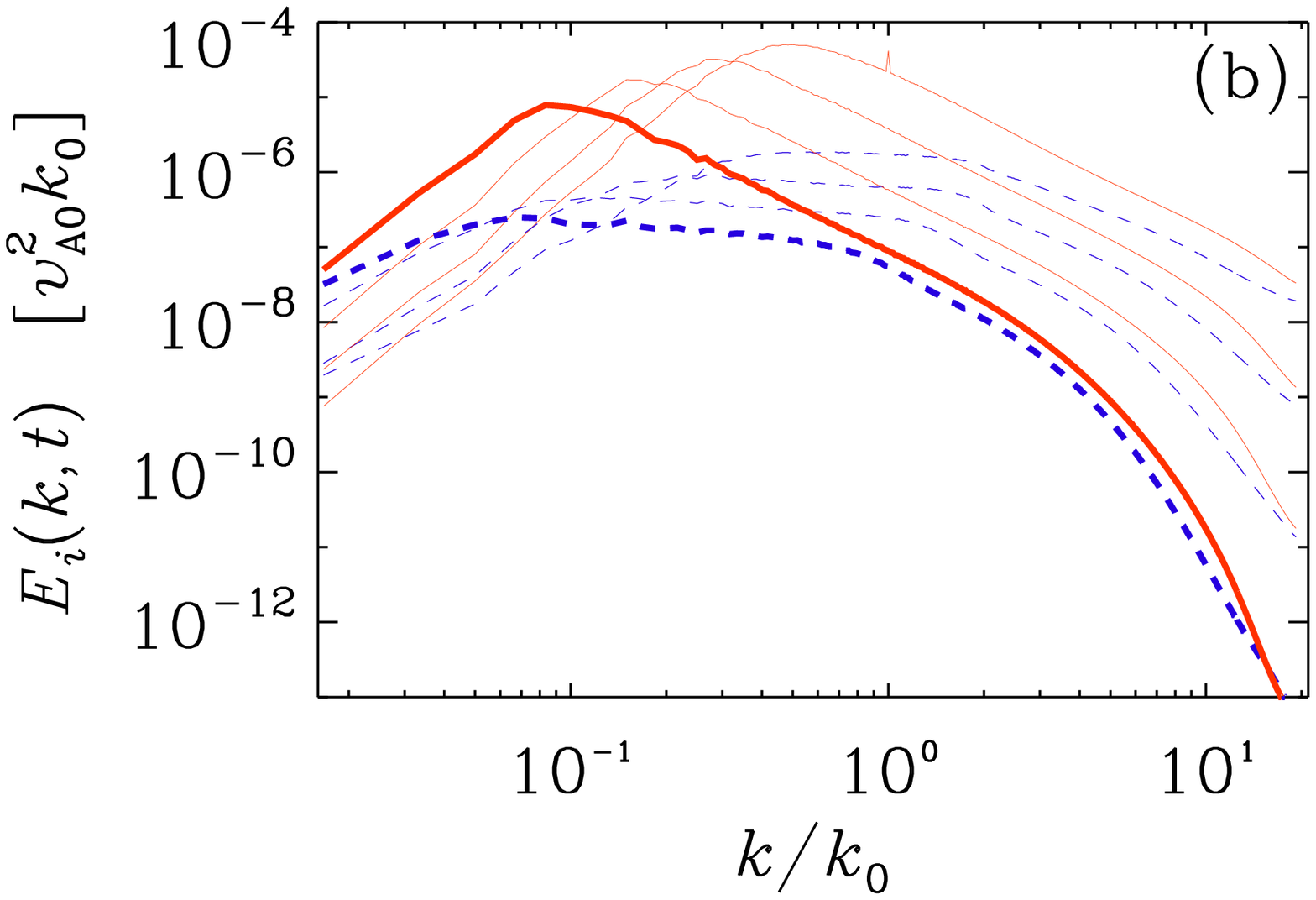}
\includegraphics[width=.32\textwidth]{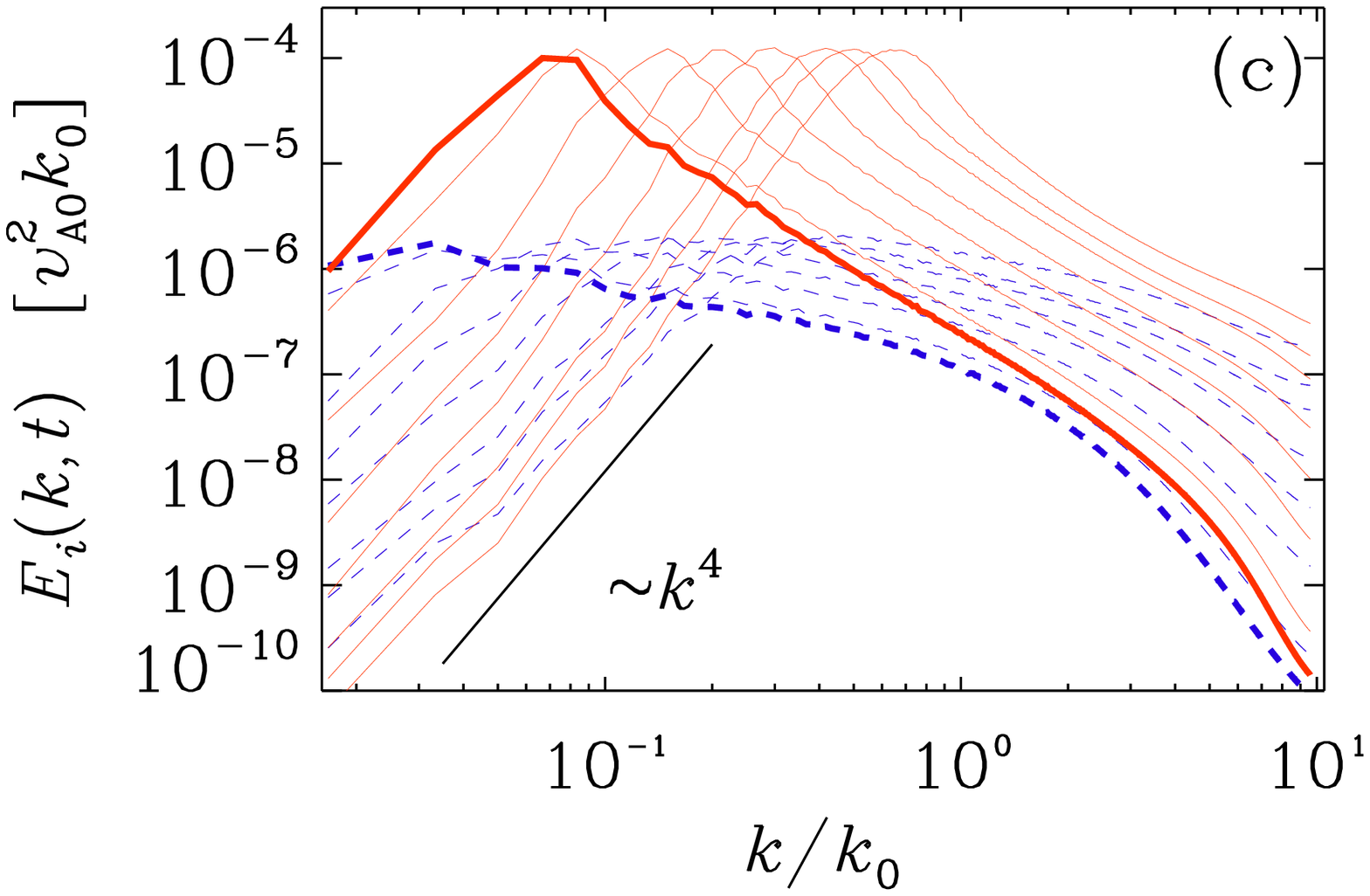}
\\
\includegraphics[width=.32\textwidth]{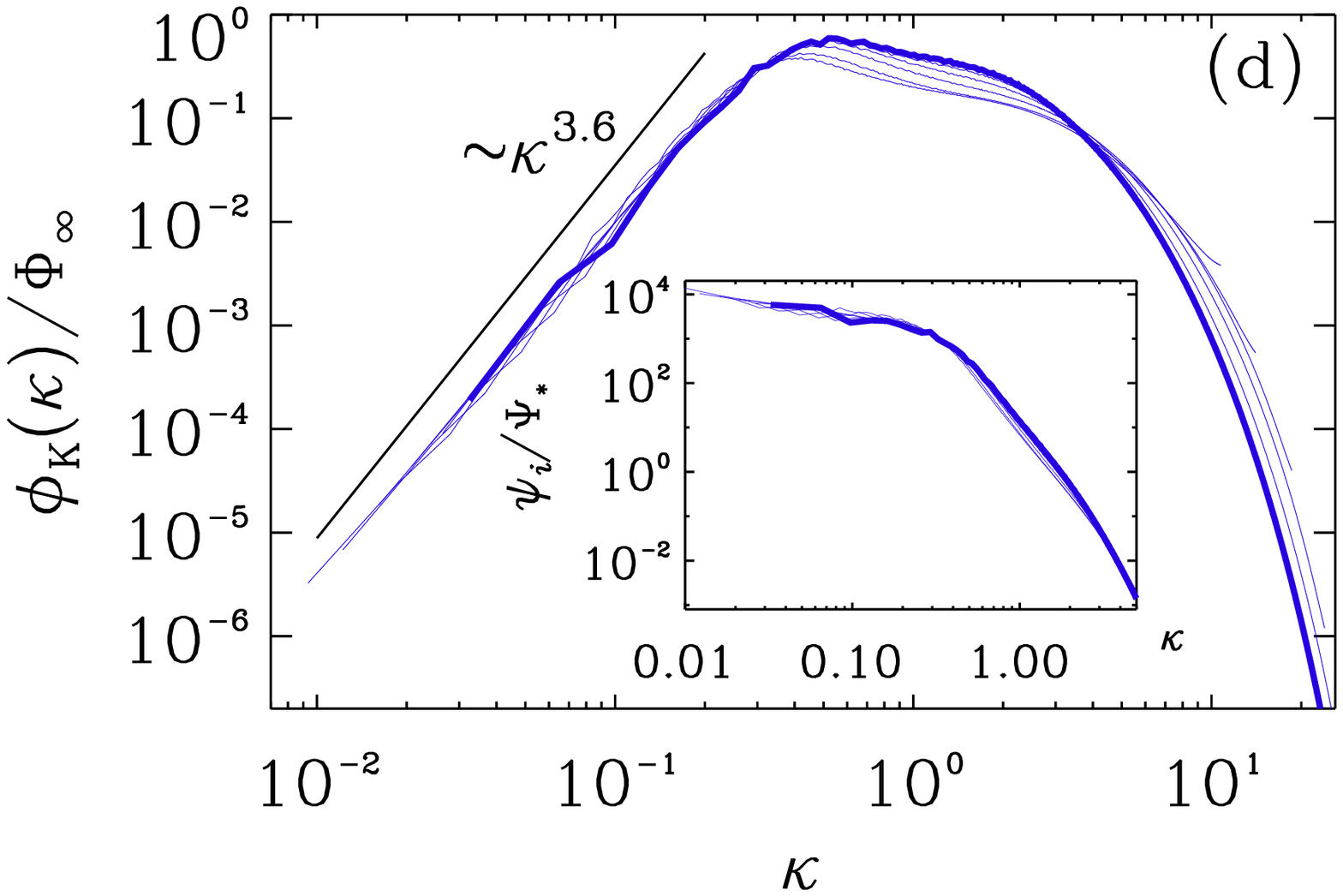}
\includegraphics[width=.32\textwidth]{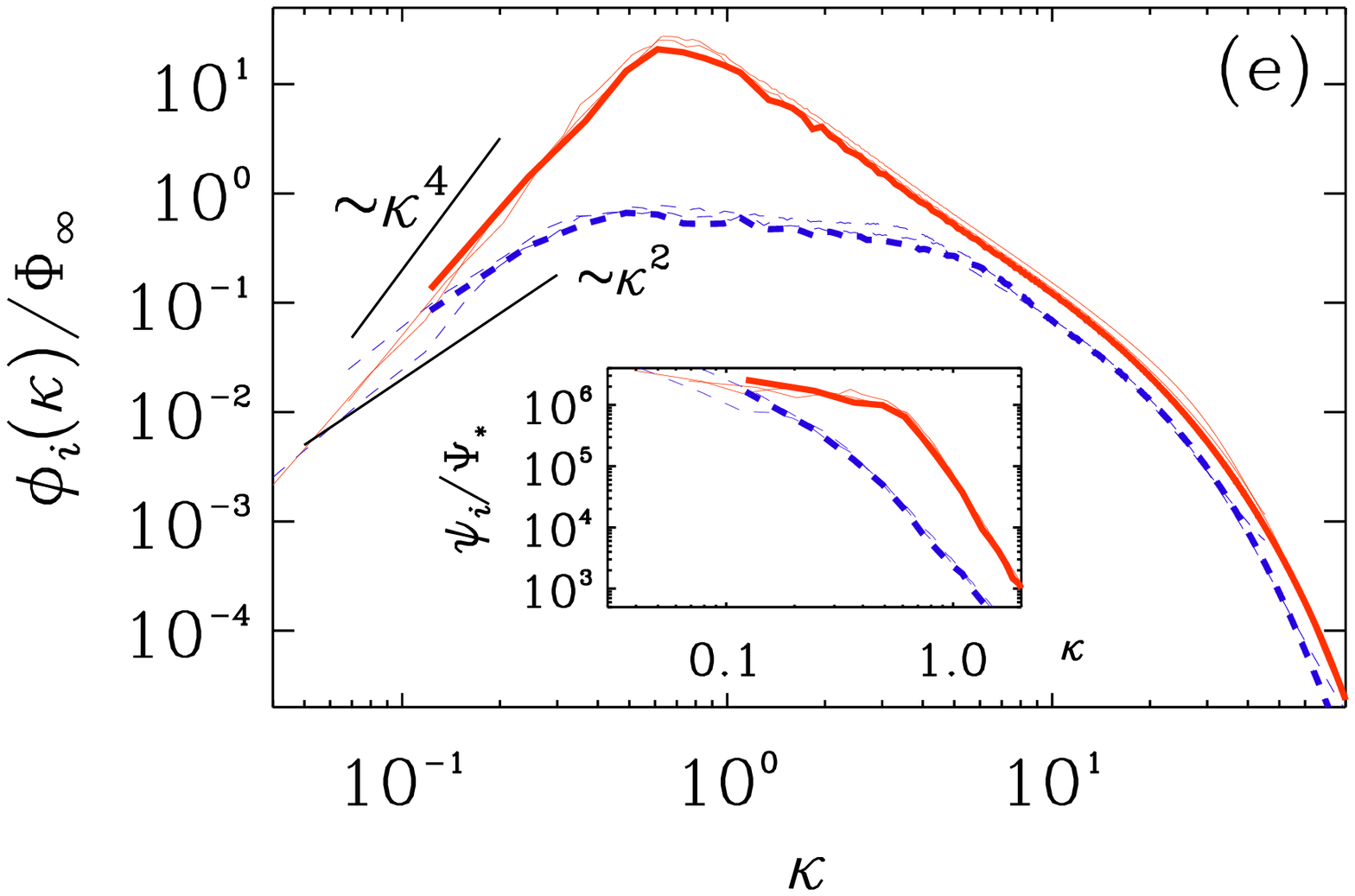}
\includegraphics[width=.32\textwidth]{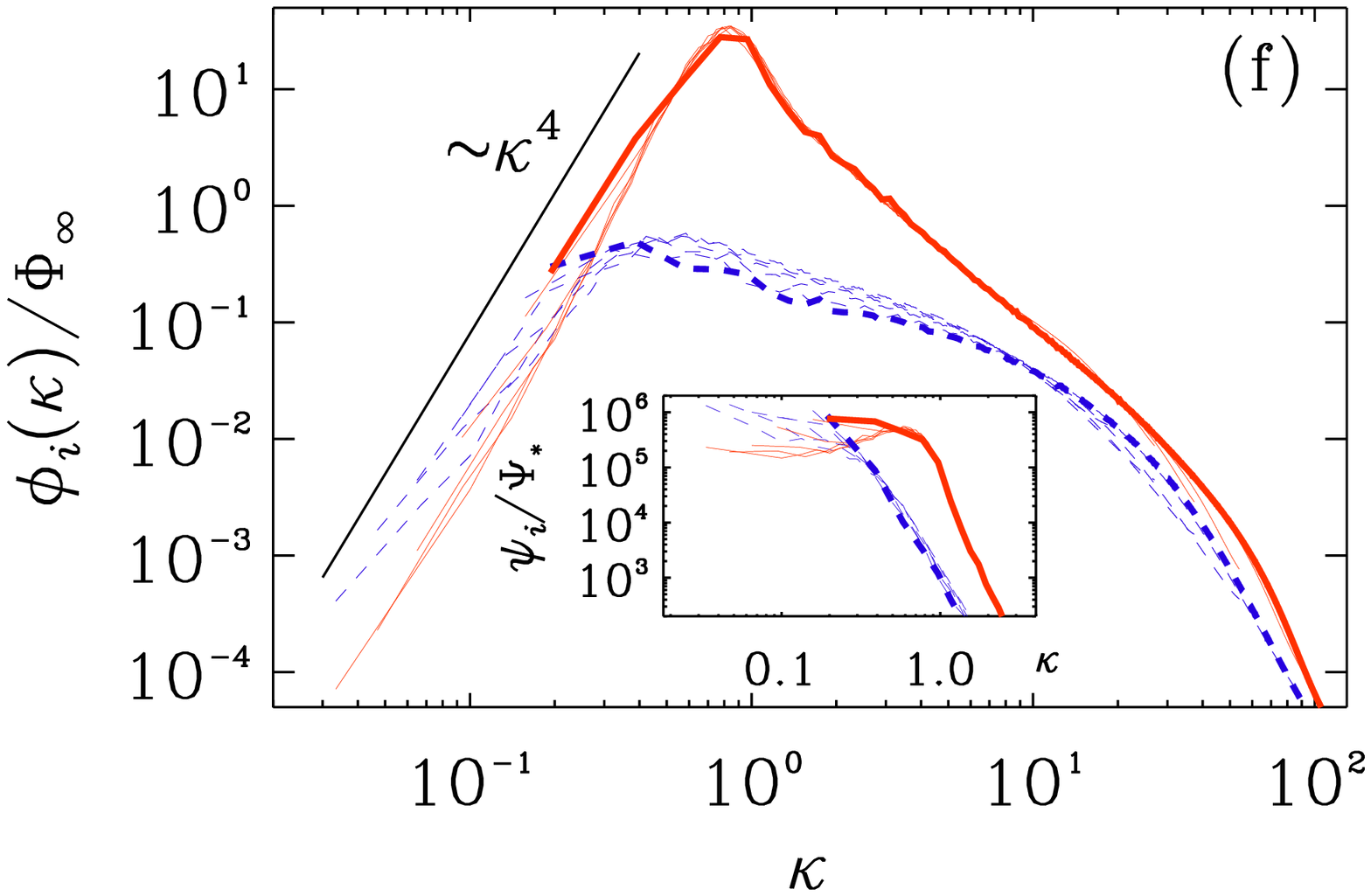}
\\
\caption[]{
$\EK(k,t)$ for different $t$ in HD DNS (a), compared with $\EM$ (solid red)
and $\EK$ (dashed blue) in MHD without helicity (b), and with (c).
Panels (d)--(f) show collapsed spectra using $\beta=3$ (d),
$\beta=1$ (e), and $\beta=0$ (f).
}\label{pkt1152_Kol1152_k4b_120k0}
\end{figure*}

In this Letter, we argue that the scaling exponent $q$ is
{\em not} primarily determined by the initial value of $\alpha$,
as suggested by \Eq{qrelation}, but by the physical processes involved.
Moreover, we relax the restriction $\psi'(0)=0$ and write instead
\EQ
E(k\xi(t),t)=\xi^{-\beta}\phi(k\xi),
\label{collapse2}
\EN
where $\xi=\xi(t)$ is computed from \Eq{xirelation2}, and $\beta$
needs to be determined empirically or theoretically.
Clearly, the initial powerlaw slope at small $k$ is no longer
an adjustable input parameter, but is fixed by the form of
$\phi=\phi(\kappa)$.
Specifically, the ``intrinsic'' slope is
$\alpha_\ast\equiv d\ln\phi/d\ln\kappa$.
Evidently, $\psi$ can be computed from $\phi$ as
$\psi(\kappa)=\xi^{\alpha-\beta}\phi(\kappa)/\kappa^\alpha$, but,
in general, $d\ln\psi/d\ln\kappa=\alpha_\ast-\alpha\neq0$ for $\kappa\to0$.

In the following, we study examples of different decay behaviors
in the diagnostic $pq$ diagram using data from DNS.
As in earlier work \cite{TKBK12}, we solve the nonideal HD and MHD
equations for an isothermal equation of state, i.e., pressure $P$ and
density $\rho$ are proportional to each other, $P=\rho\cs^2$, where
$\cs=\const$ is the sound speed.
The kinematic viscosity $\nu$ is characterized by the Reynolds
number, $\Rey=\urms\xi/\nu$, with $\urms=(2{\cal E})^{1/2}$ and the
magnetic diffusivity $\eta$ is characterized by the magnetic Prandtl
number $\Pm=\nu/\eta$.
The governing equations are solved using the {\sc Pencil Code} \cite{PC,suppl}.
The resolution is either $1152^3$ or $2304^3$ meshpoints.
The Mach number $\urms/\cs$ is always below unity, so
compressibility effects are weak.

We first consider cases that have $\alpha=4$ for the initial spectral
slopes of $\EK$ or $\EM$.
We consider (i) HD decay, (ii) nonhelical MHD decay,
and (iii) helical MHD decay.
In cases (ii) and (iii), the magnetic energy also drives kinetic
energy through the Lorentz force.
The particular simulation of case~(ii) was already presented in
Ref.~\cite{BKT15}, where inverse transfer to smaller wavenumbers was
found in the absence of magnetic helicity using high-resolution DNS.
Case~(iii) leads to standard inverse transfer \cite{PFL76,BM99,CHB01,BJ04}.
The resulting spectra are plotted in \Figssp{pkt1152_Kol1152_k4b_120k0}{a}{c},
where we show energy spectra for cases (i)--(iii) at different times.
The values of $\Rey$ at half time are roughly $100$, $230$, and $300$,
respectively.

In \Figssp{pkt1152_Kol1152_k4b_120k0}{d}{f} we compare with suitably
compensated spectra.
We compensate for the shift in $k$ by plotting $E(k,t)$
against $k\xi(t)$.
The peak in each spectrum, which is approximately at
$k=\xi^{-1}$, has then always the same position on the abscissa.
Furthermore, to compensate for the decay in energy, we multiply $E$
by $\xi^\beta$ with some exponent $\beta$ such that the
compensated spectra collapse onto a single function
$\phi(k\xi(t))\approx\xi^\beta E(k\xi(t),t)$.
In terms of the energy ${\cal E}(t)\equiv\int E(k,t)\,dk$, the function
$\Phi=\xi^{\beta+1}{\cal E}_{\rm K}$ is asymptotically constant,
$\Phi(t)\to\Phi_\infty$, and has the same dimension as $\phi$,
so we plot the nondimensional ratio $\phi/\Phi_\infty$.
The function $\psi(\kappa)$ is shown as an inset and normalized
by $\Psi_\ast\equiv\xi^{\alpha-\beta}\Phi$ at the last time.

\begin{figure*}[t!]
\includegraphics[width=\textwidth]{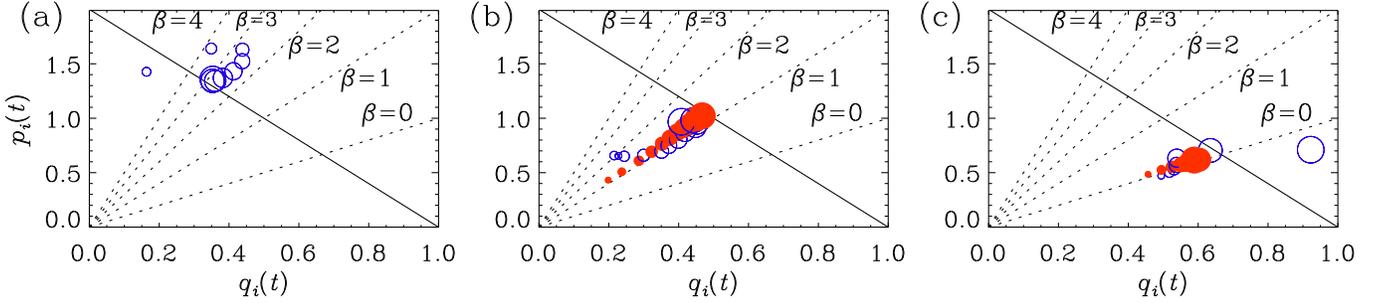}
\caption[]{
$pq$ diagrams for cases (i)--(iii).
Open (closed) symbols correspond to $i={\rm K}$ (${\rm M}$) and their
sizes increase with time.
}\label{pqcomp}
\end{figure*}

Let us now consider solutions (i)--(iii) in the $pq$ diagram;
see \Figssp{pqcomp}{a}{c}.
These are compatible with independently computed $\beta q$ diagrams \cite{suppl}.
To study the relation between the exponents $\beta$ and $q$, we
make use of Olesen's scaling arguments and that $\phi$ is invariant
under rescaling, to show from \Eq{collapse2} that $\beta+3-2/q=0$, i.e.,
\EQ
\beta=2/q-3,
\label{betarelation}
\EN
or $q=2/(3+\beta)$.
This is formally equivalent to Olesen's relation \eq{qrelation},
but with $\alpha$ being replaced by $\beta$.
Moreover, unlike the exponent $\alpha$ in \Eq{scalinglaw}, the exponent
$\beta$ in \Eq{collapse2} bears no relation with the initial spectral slope,
except for certain cases discussed below.
The temporal decay of kinetic and magnetic energies
follows power laws ${\cal E}_i(t)\sim t^{-p_i}$ for
$i={\rm K}$ or ${\rm M}$.
The exponents are obtained by integrating over $k$,
${\cal E}(t)=\xi^{-(\beta+1)}\int\phi\,d(k\xi)\propto t^{-p}$,
and since $\xi\propto t^q$, this yields
\EQ
p=(1+\beta)\,q.
\label{pqbeta}
\EN
Thus, in a $pq$ diagram, a certain value of $\beta$ corresponds to a
line $p(t)\propto q(t)$ with the slope $\beta+1$.
Furthermore, inserting \Eq{betarelation} yields the line $p=2(1-q)$.
We call this the self-similarity line.

\begin{figure*}[t!]
\includegraphics[width=.32\textwidth]{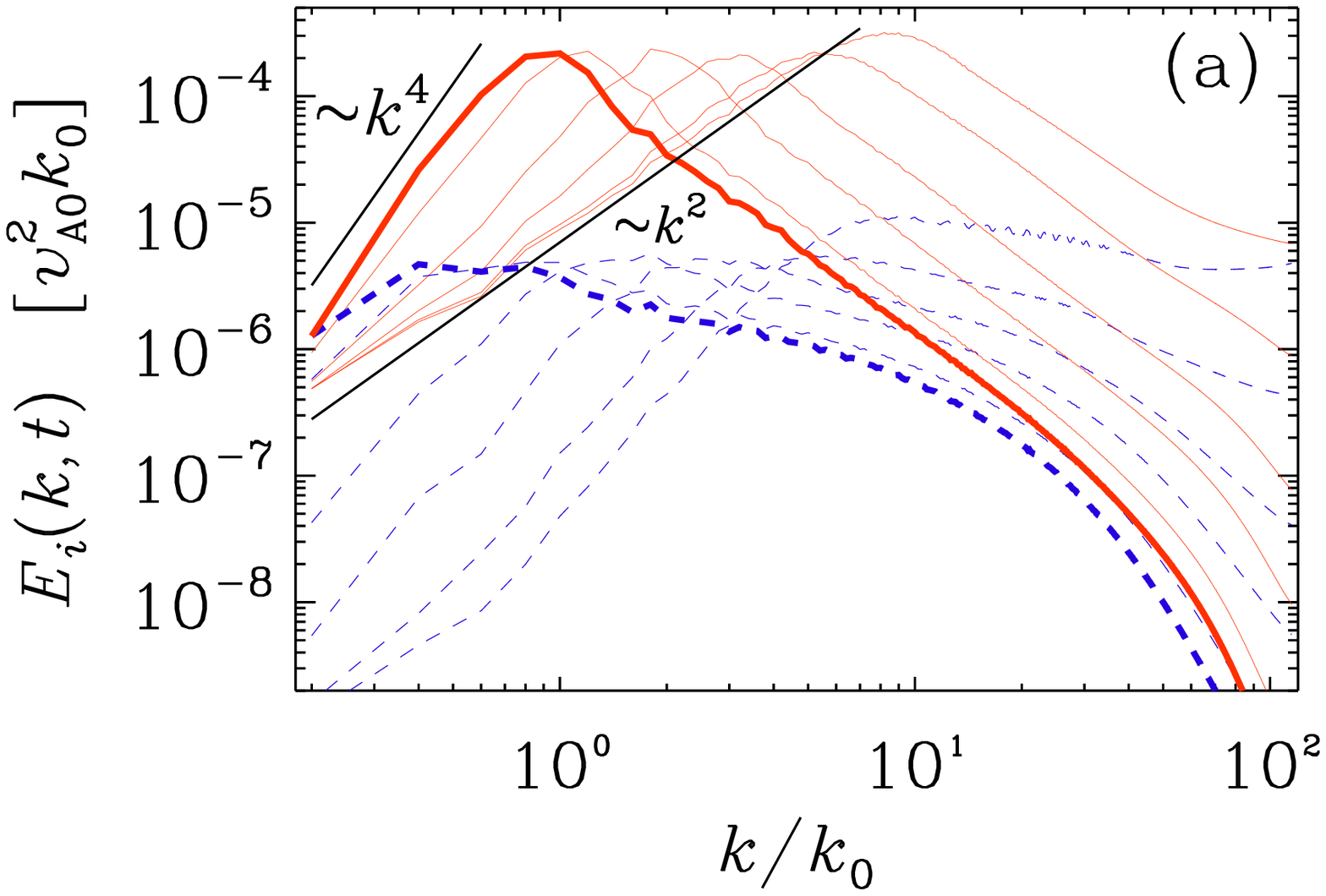}
\includegraphics[width=.32\textwidth]{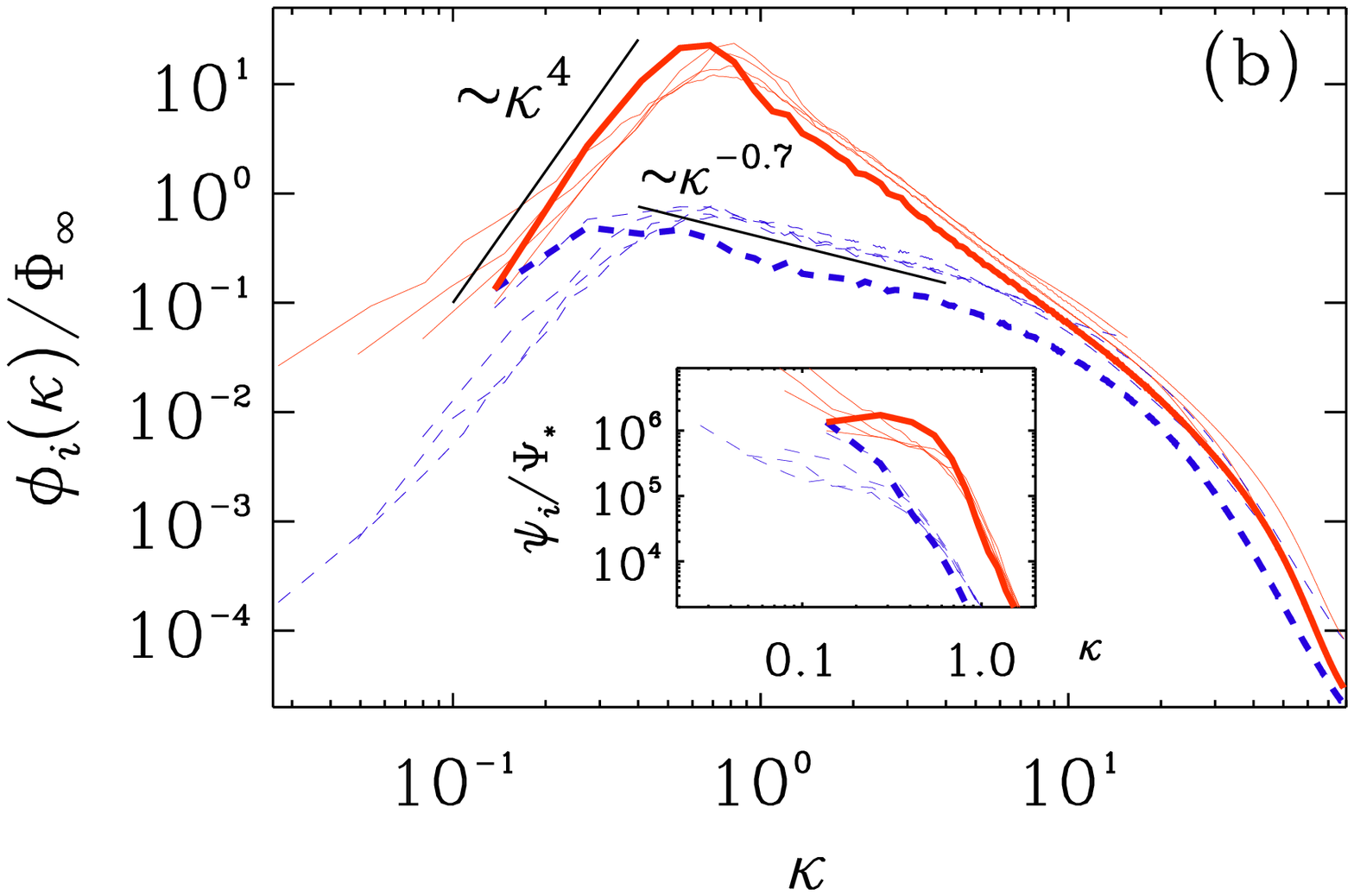}
\includegraphics[width=.32\textwidth]{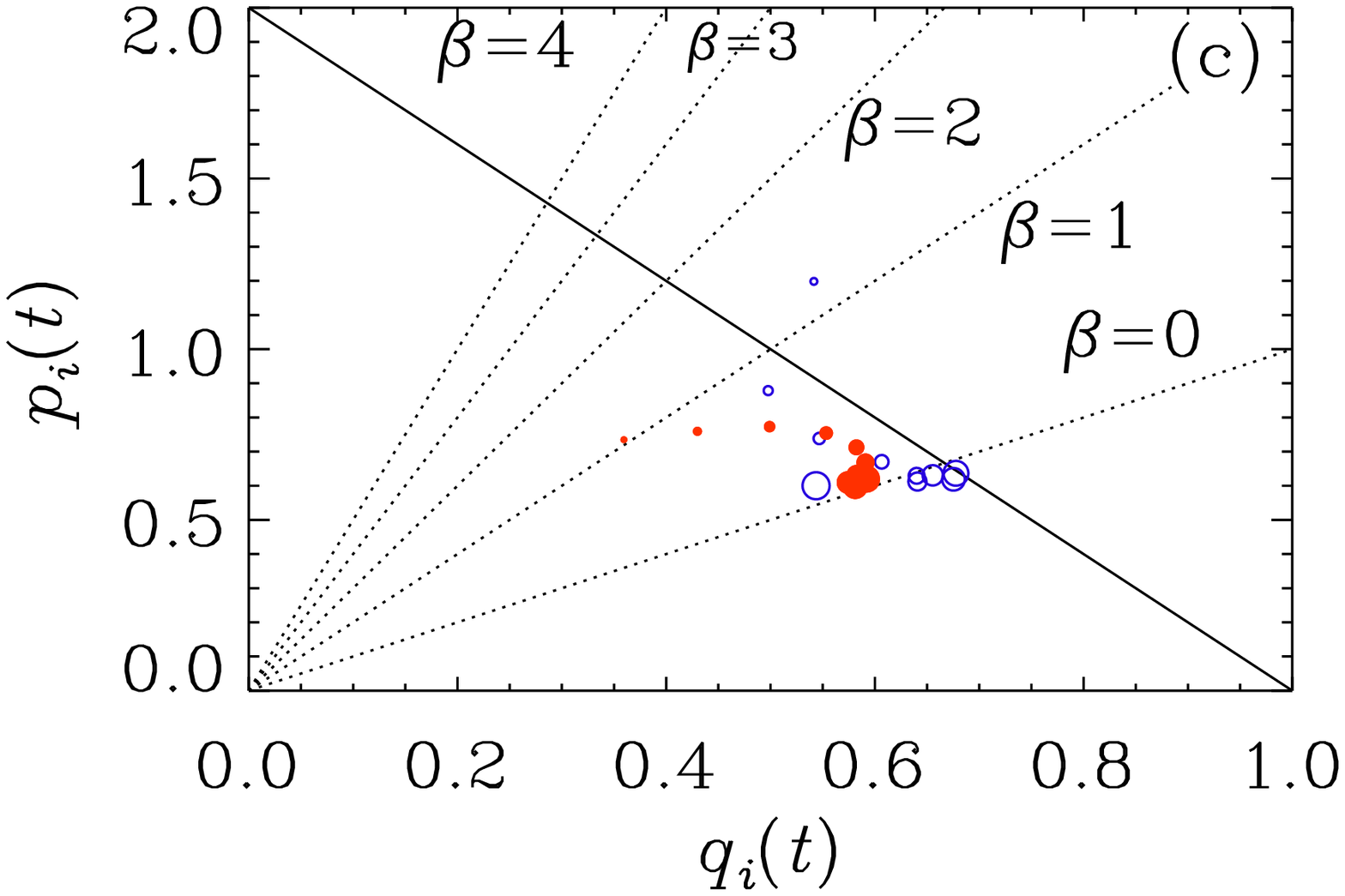}
\\
\includegraphics[width=.32\textwidth]{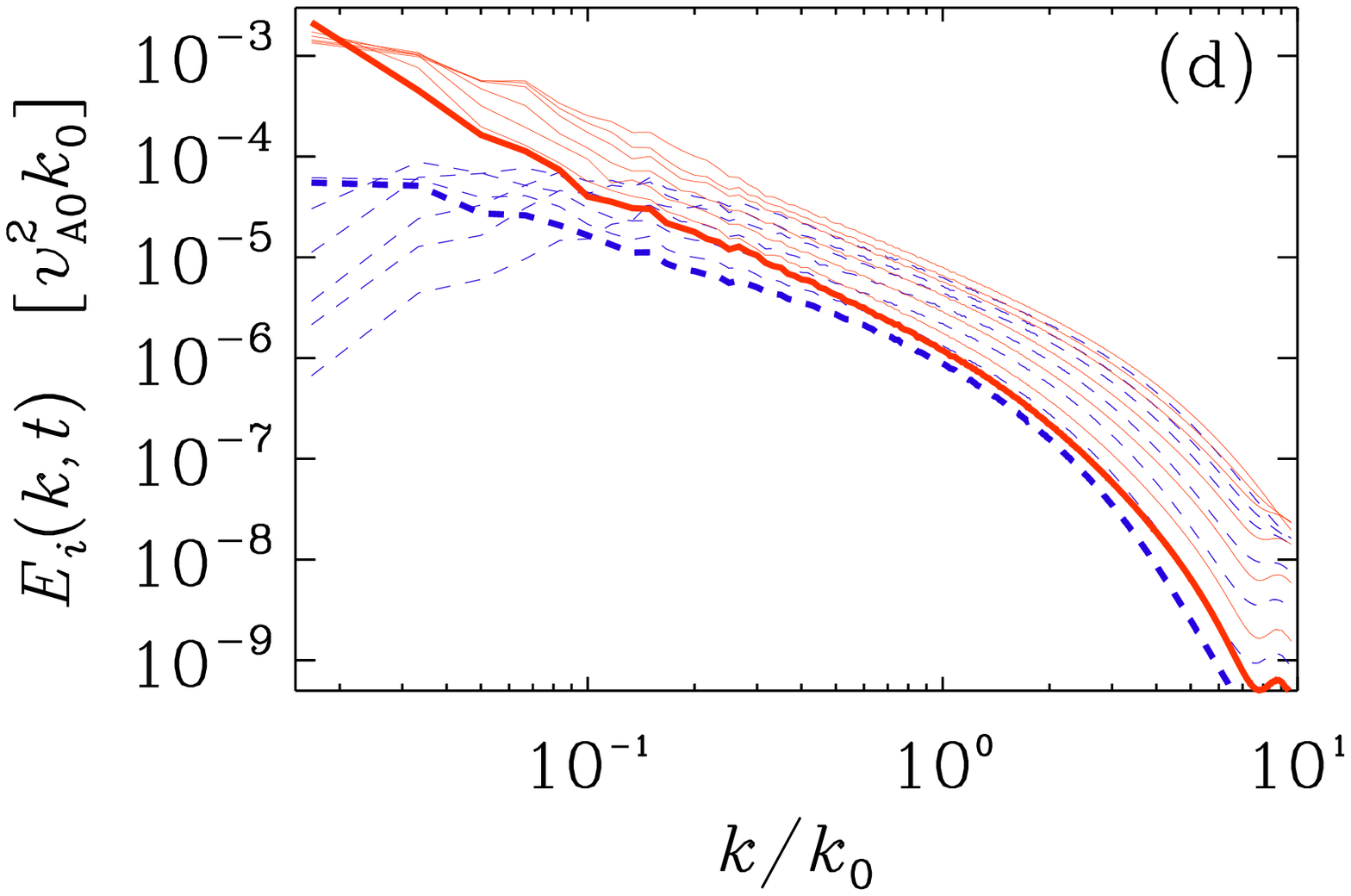}
\includegraphics[width=.32\textwidth]{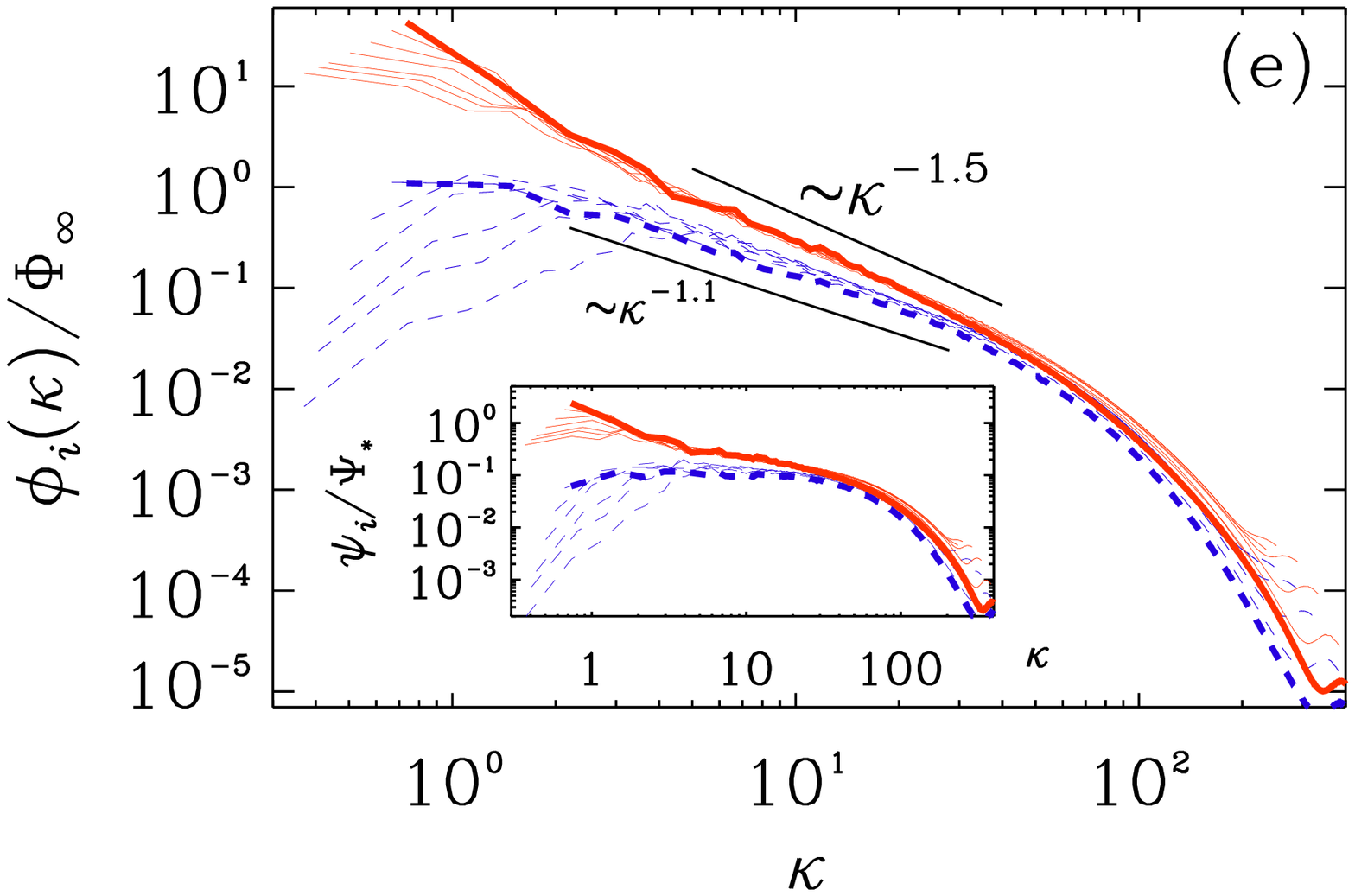}
\includegraphics[width=.32\textwidth]{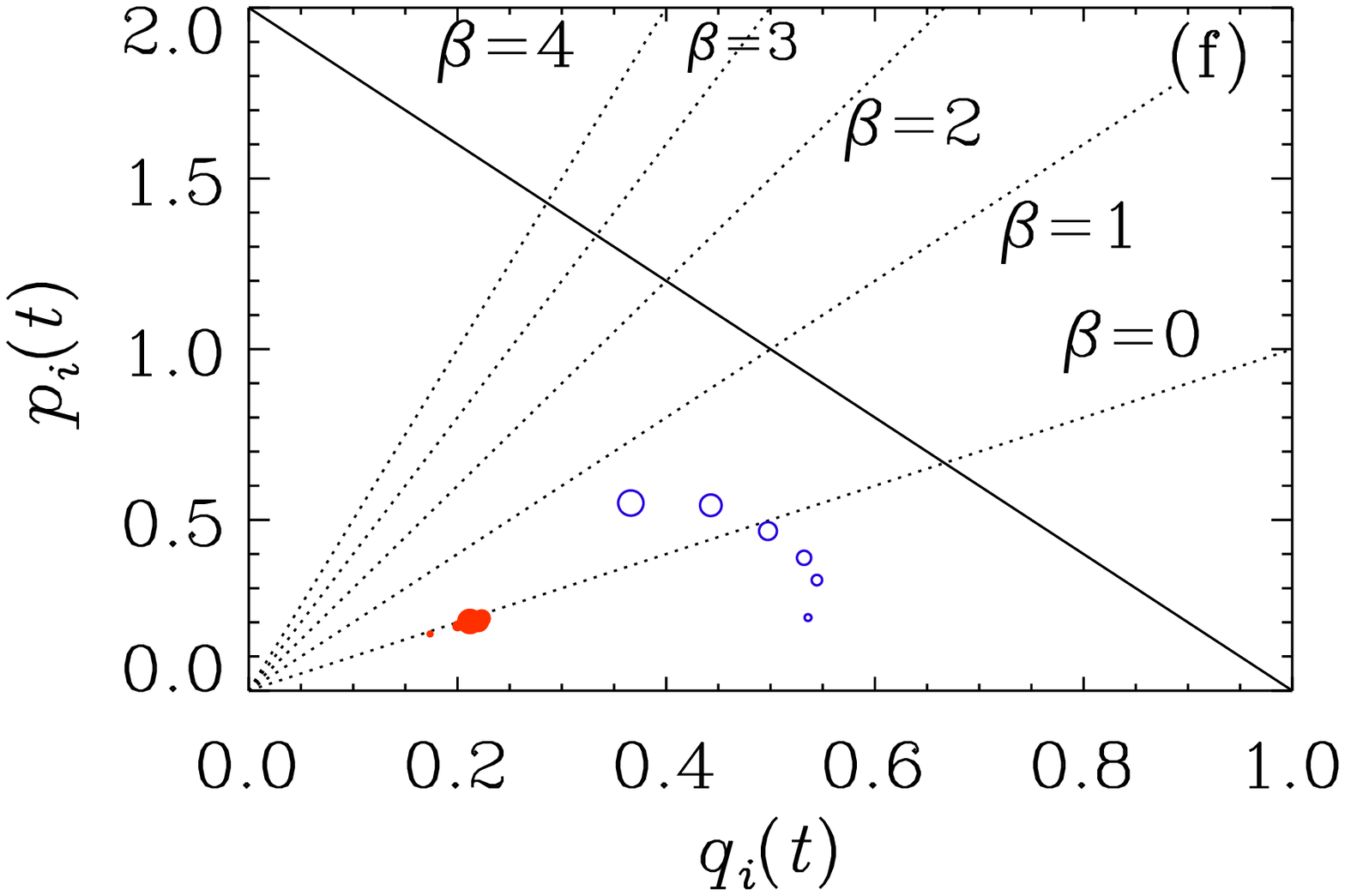}
\caption[]{
$\EM$ (solid) and $\EK$ (dashed) in MHD with fractional helicity and $\alpha=2$ (a),
as well as full helicity and $\alpha=-1$ (d), together with compensated spectra (b,e)
and the $pq$ diagrams (c,f).
}\label{pkt1152_H1152k2b_sig01}
\end{figure*}

\begin{figure*}[t!]
\includegraphics[width=.32\textwidth]{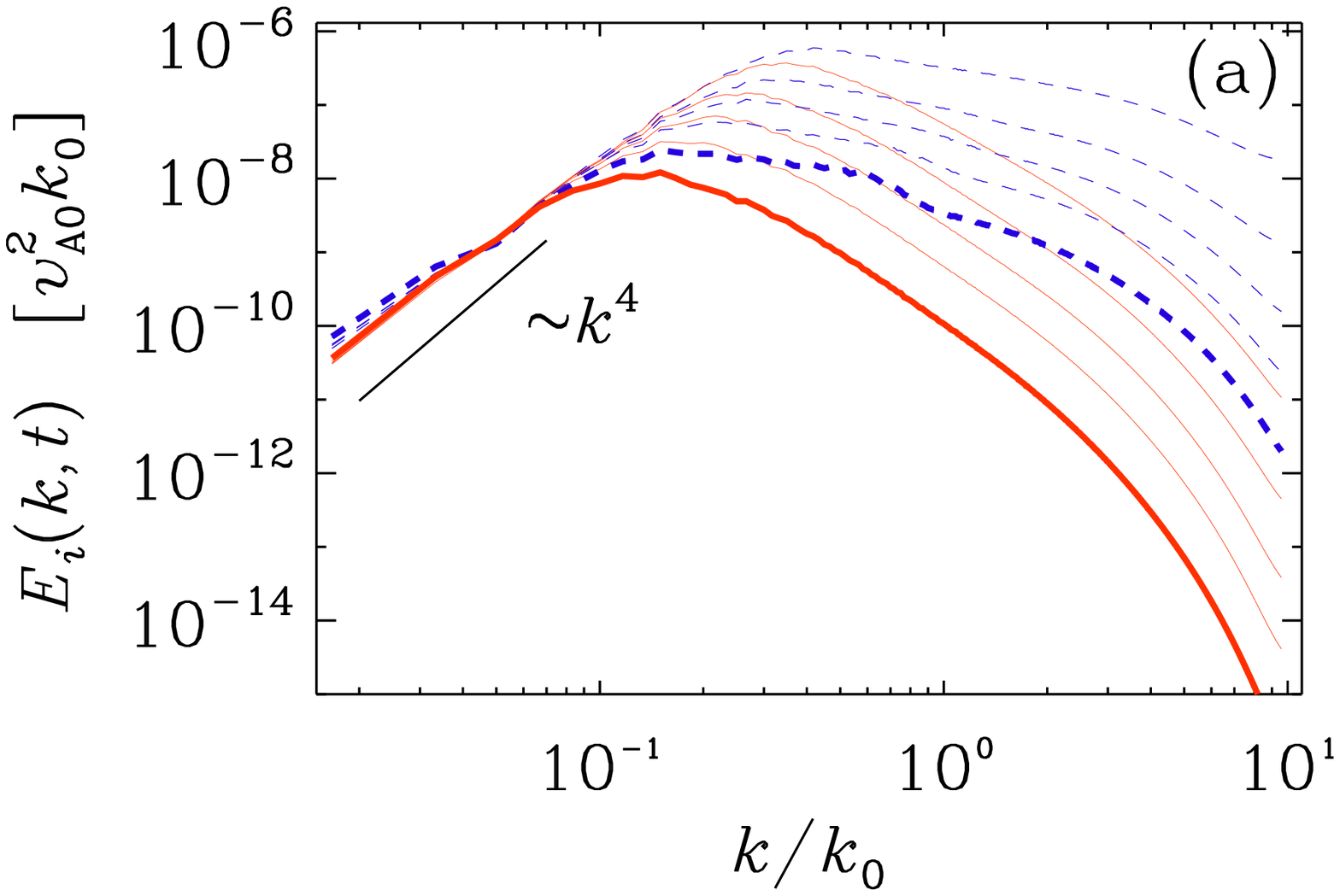}
\includegraphics[width=.32\textwidth]{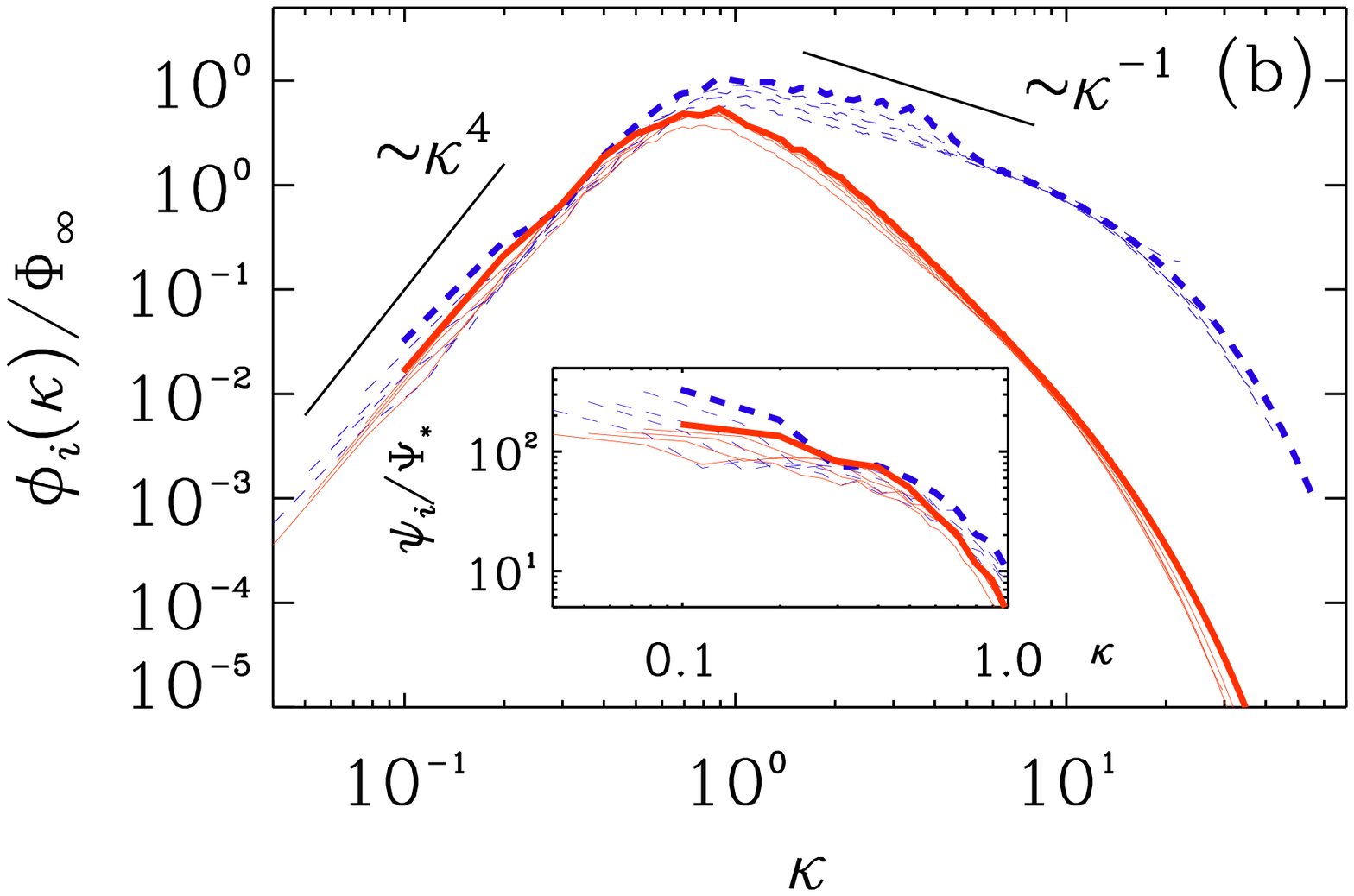}
\includegraphics[width=.32\textwidth]{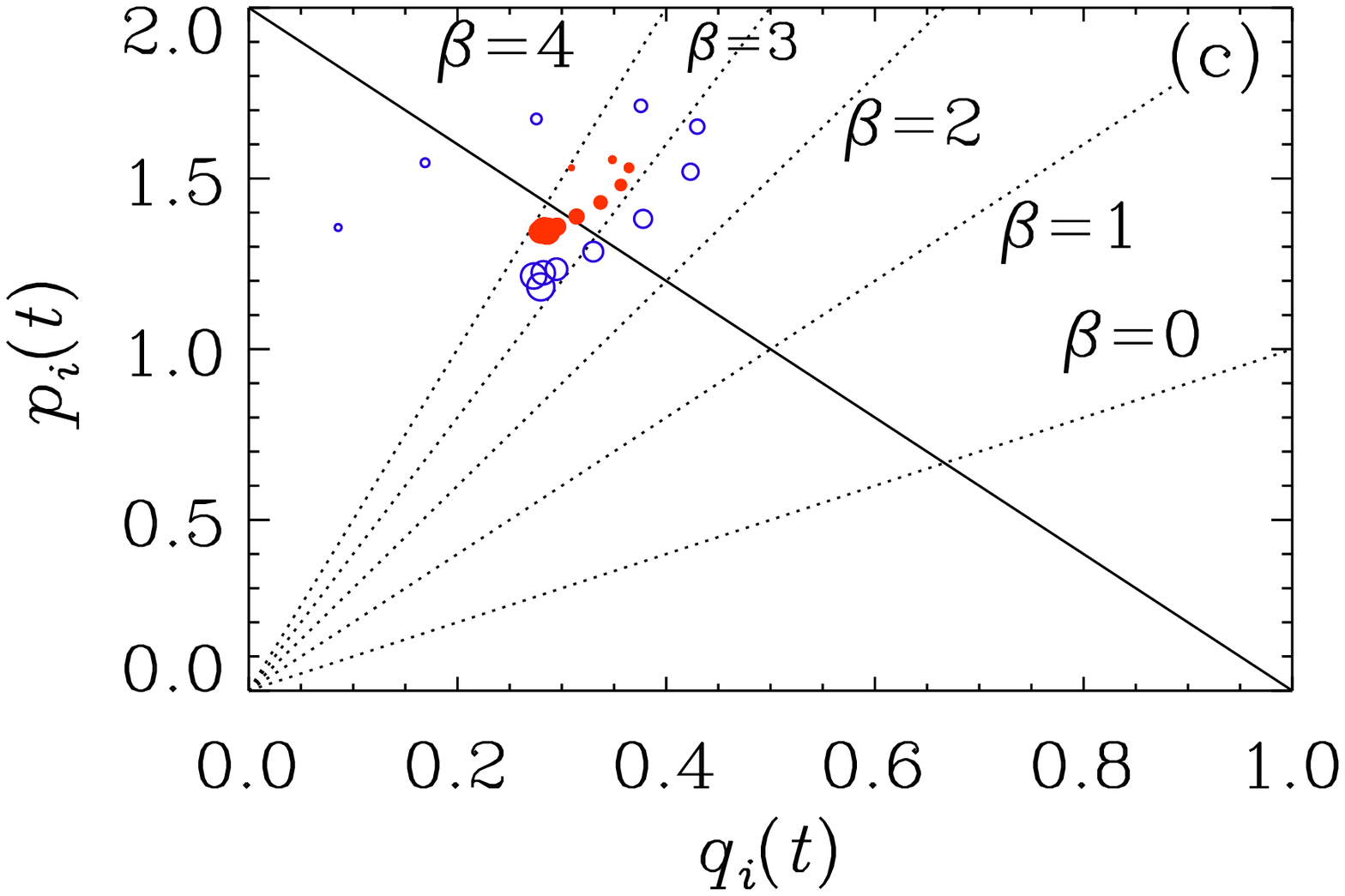}
\includegraphics[width=.32\textwidth]{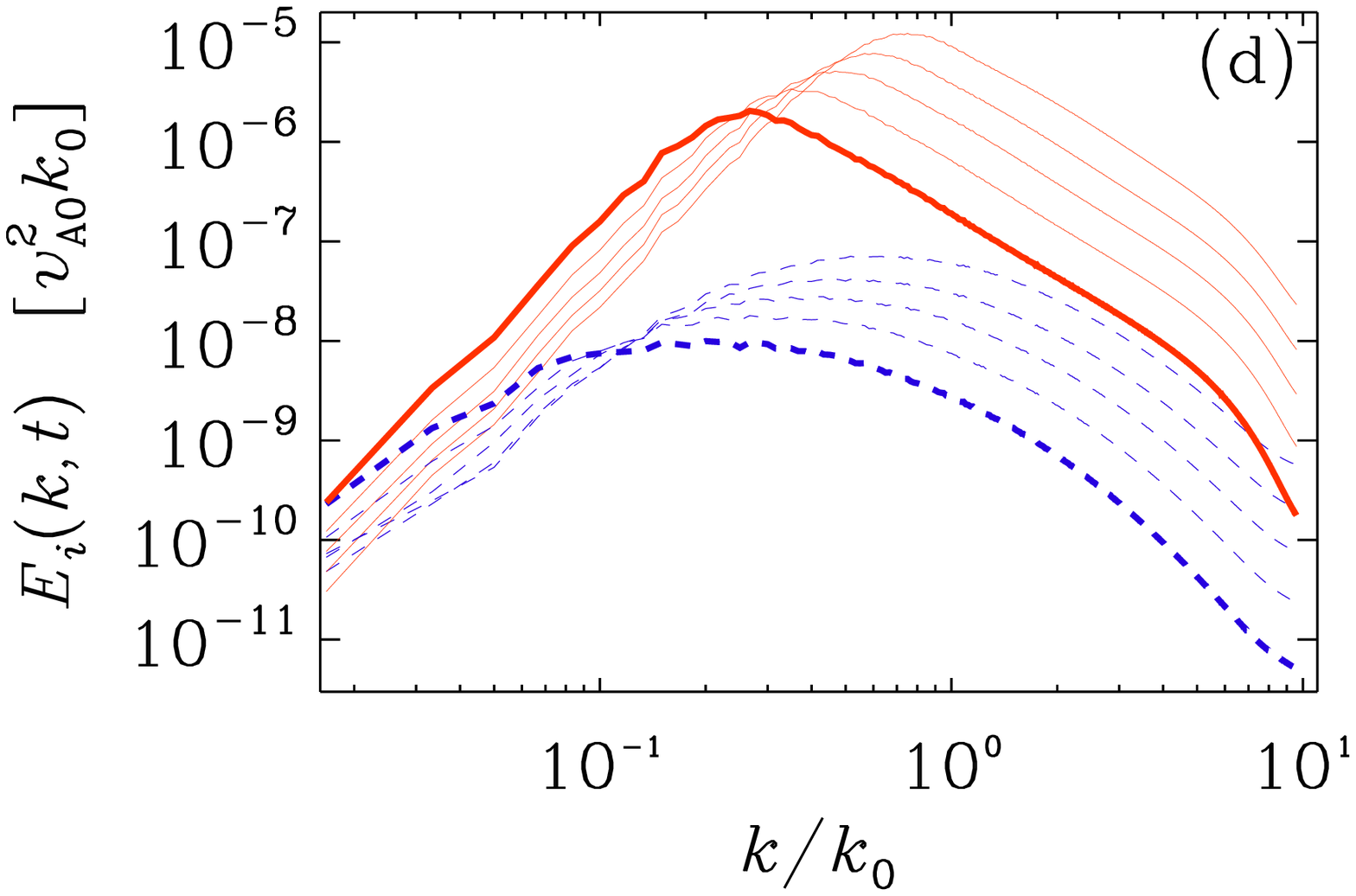}
\includegraphics[width=.32\textwidth]{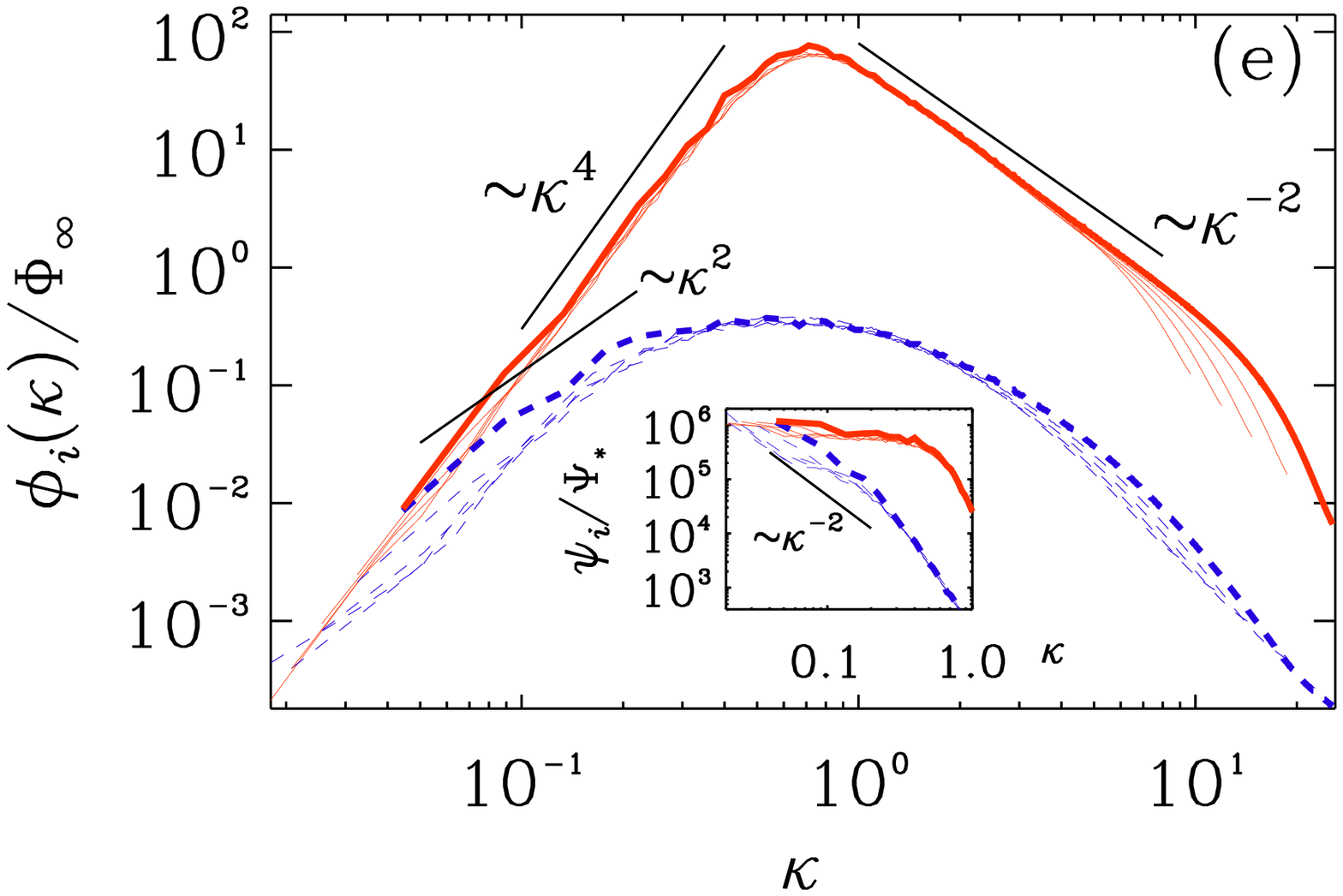}
\includegraphics[width=.32\textwidth]{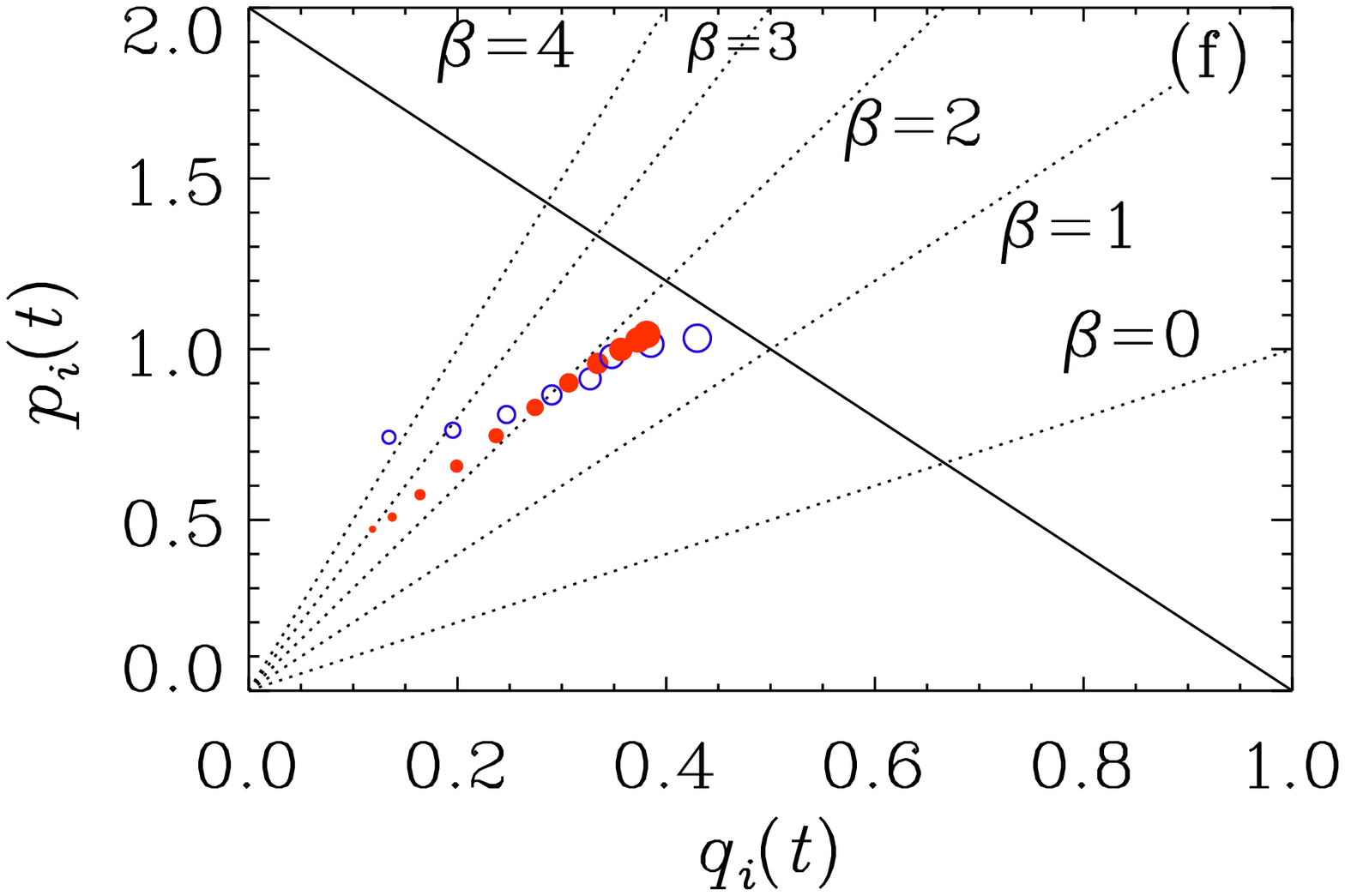}
\caption[]{
Similar to \Fig{pkt1152_H1152k2b_sig01}, but for nonhelical MHD with
$\Pm=0.01$ (a) and $\Pm=100$ (d), together with compensated spectra (b,e)
and the $pq$ diagrams (c,f).
}\label{Pm100}
\end{figure*}

\begin{table}[b!]\caption{
Scaling exponents and relation to physical invariants and their dimensions.
}\vspace{12pt}\centerline{\begin{tabular}{rrrll}
$\beta$ & $p\quad\quad$ & $q\quad\quad\;$ & $\quad$inv. & dim. \\
\hline
4 & $\quad10/7\approx1.43$ & $\quad2/7\approx0.286$ &
$\quad{\cal L}$ & $[x]^7[t]^{-2}$ \\
3 &   $8/6\approx1.33$     & $2/6\approx0.333$      & \\
2 &   $6/5=1.20$           & $2/5=0.400$            & \\
1 &   $4/4=1.00$           & $2/4=0.500$            &
$\quad\bra{\AAA_{\rm 2D}^2}$ & $[x]^4[t]^{-2}$ \\
0 &   $2/3\approx0.67$     & $2/3\approx0.667$      &
$\quad\bra{\AAA\cdot\BB}$  & $[x]^3[t]^{-2}$   \\
$-1$& $0/2=0.00$           & $2/1=1.000$      \\
\label{TSum}\end{tabular}}
\end{table}

The exponents $\beta$, $p$, and $q$ are roughly consistent
with those expected based on the dimensions of potentially conserved
quantities such as the Loitsiansky integral \cite{Dav10},
${\cal L}=\int\rr^2\bra{\uu(\xx)\cdot\uu(\xx+\rr)}\,d\rr\propto\ell^5 u_\ell^2$,
with typical velocity $u_\ell$ on scale $\ell$,
the magnetic helicity, $\bra{\AAA\cdot\BB}$, where
$\BB=\nab\times\AAA$ is the magnetic field in terms of
the vector potential $\AAA$, and the mean squared vector potential,
$\bra{\AAA^2}$, which is conserved in two-dimensions (2D); see \Tab{TSum}.

In the HD case (i), the solution approaches the
$\beta=3$ line and then settles on the self-similarity line at
$q\approx1/3$; see \Figp{pqcomp}{a}.
This decay behavior departs from what would be expected if the
Loitsiansky integral were conserved, i.e., $q=2/7$ and $\beta=4$.
A slower decay law with $p=6/5$, corresponding to $q=2/5$ and $\beta=2$
has been favored by Saffman \cite{Saf67}, while experiments and
simulations suggest $p=5/4$ \cite{Kang,HB04}.

In case (ii), the solution evolves along $\beta=1$ toward $q=1/2$;
see \Figsp{pqcomp}{b}{e}.
This is compatible with the conservation of $\bra{\AAA_{\rm 2D}^2}$,
where $\AAA_{\rm 2D}$ is the component of $\AAA$ which describes the
2D magnetic field in the plane perpendicular to the local
intermediate eigenvector of the rate-of-strain matrix $\SSSS$; see the
supplemental material of \cite{BKT15} for details, and also \cite{Ole15}.
The motivation for applying 2D arguments to 3D comes from the fact that
for sufficiently strong magnetic fields the dynamics tends to become
locally 2D in the plane perpendicular to the local field.
This allows one to compute $\AAA$ in a gauge that projects out
contributions perpendicular to the intermediate eigenvector of $\SSSS$.

In case (iii) the solution evolves along $\beta=0$
toward $q=2/3$; see \Figsp{pqcomp}{c}{f}.
This means that the spectrum shifts just in $k$, while the
amplitude of $\EM$ does not change, as can be seen from
\Figp{pkt1152_Kol1152_k4b_120k0}{c}.
This is consistent with the invariance of $\bra{\AAA\cdot\BB}$;
see Ref.~\cite{BM99}.

Next, we investigate cases with $\alpha<4$.
In the helical case with $\alpha=2$ we see that the
subinertial range spectrum quickly steepens and approaches
$\alpha_\ast=4\neq\alpha$; see \Figssp{pkt1152_H1152k2b_sig01}{a}{c}.
For $\alpha=-1$, which is a scale-invariant spectrum, the
spectral energy remains nearly unchanged at small $k$, but the
magnetic energy still decays due to decay at all higher $k$; see
\Figssp{pkt1152_H1152k2b_sig01}{d}{f}.
The values of $\pM$ and $\qM$ are rather small ($\approx0.2$), but
the spectra can still be collapsed onto each other with $\beta=0$;
see \Figp{pkt1152_H1152k2b_sig01}{e}.

The examples discussed above demonstrate that in general
$\beta\neq\alpha\neq\alpha_\ast$,
i.e., the self-similarity parameter $\beta$ is not determined by the initial
power spectrum but rather by the different physical processes involved.
In helical MHD, we always find $\alpha_\ast=4$ together with $\beta=0$.
For nonhelical MHD with $\alpha=4$ and $E_{\rm K}\propto k^2$,
we find $\beta=1$, while in HD with $\alpha=4$, we find $\beta=3$.
In agreement with earlier work \cite{YHB04}, the following exceptions
can be identified: in HD with $1\leq\alpha\leq3$
and in nonhelical MHD with $1\leq\alpha\leq$ we find $\beta=\alpha$
\cite{suppl}.
The only case where $\alpha=\beta=4$ has been found is when the magnetic
Prandtl number $\Pm\equiv\nu/\eta$ is small; see \Figsp{Pm100}{a}{c} for $\Pm=0.01$.
Here, the conservation of ${\cal L}$ may actually apply \cite{Dav10}.
For $\Pm\equiv\nu/\eta\gg1$, on the other hand, we find $\beta=2$ scaling, even
though $\alpha=4$; see \Figsp{Pm100}{d}{f}.

In conclusion, the present work has revealed robust properties
of the scaling exponent $\beta$ governing the time-dependence
of the energy spectrum $E(k,t)$ through $\xi^\beta\phi(k\xi)$
with a time-independent scaling function $\phi$ and a time-dependent
integral scale $\xi(t)$.
The helical case is particularly robust in that any point in the $pq$
plane evolves along the $\beta=0$ line ($p=q$) toward the point $p=q=2/3$.
Furthermore, if the initial spectrum has $\alpha=2$, it first steepens to
$\alpha=4$ and then follows the same decay as with an initial $\alpha=4$.
Moreover, for a scale-invariant spectrum with $\alpha=-1$, we again find
$\beta=0$, i.e., the same as for $\alpha=2$ and 4, but now with
$\pM\approx\qM\approx0.2$; see \Figp{pkt1152_H1152k2b_sig01}{f}.
In the fractionally helical case, points in the $pq$ plane
evolve toward the $\beta=0$ line and, for $\alpha\ge2$,
later toward $\pM=\qM=2/3$.

Our results have consequences for two types of cosmological initial
magnetic fields:
causal ones with $\EM\propto k^4$ will always be accompanied by a shallower
kinetic energy spectrum $\EK\propto k^2$, thus favoring inverse transfer
\cite{BKT15,KTBN13}, while a scale-invariant inflation-generated helical
field exhibits self-similarity with $\beta=0$ in the same way as for other
initial slopes, but now with $p=q\approx0.2$ instead of $2/3$.
For decaying wind tunnel turbulence, Loitsiansky scaling is ruled out in
favor of Saffman scaling, provided $\alpha=2$.
No inverse transfer is possible in HD, even if $\alpha=4$,
contrary to earlier claims \cite{Ole97}.
The experimental realization of initial conditions with $\alpha\neq2$
could be challenging for wind tunnels, but may well be possible in
plasma experiments \cite{Forest15}.

\acknowledgements
\vspace{3mm}

We thank Andrey Beresnyak, Leonardo Campanelli, Ruth Durrer, Alexander Tevzadze,
and Tanmay Vachaspati for useful discussions.
Support through the NSF Astrophysics and Astronomy Grant Program
(grants 1615940 \& 1615100), the Research Council of Norway (FRINATEK grant 231444),
the Swiss NSF SCOPES (grant IZ7370-152581), and the Georgian Shota Rustaveli NSF
(grant FR/264/6-350/14) are gratefully acknowledged.
We acknowledge the allocation of computing resources provided by the
Swedish National Allocations Committee at the Center for Parallel
Computers at the Royal Institute of Technology in Stockholm.
This work utilized the Janus supercomputer, which is supported by the
National Science Foundation (award number CNS-0821794), the University
of Colorado Boulder, the University of Colorado Denver, and the National
Center for Atmospheric Research. The Janus supercomputer is operated by
the University of Colorado Boulder.


\newpage
\LARGE
\noindent
Supplemental Material

\large
\vspace{1mm}
\noindent
to ``Classes of hydrodynamic and magnetohydrodynamic turbulent decay''
(arXiv:1607.01360)

\normalsize
\vspace{1mm}
\noindent
by A.\ Brandenburg \& T.\ Kahniashvili


\section{Effect of phase errors}

By default, the {\sc Pencil Code} uses sixth order accurate finite
difference representations for the first and second derivatives.
A low spatial order of the scheme implies that at high wavenumbers
the magnitude of the numerical derivative is reduced, leading to lower
advection speeds of the high wavenumber Fourier components.
This is generally referred to as phase error.
Thus, for an advected tophat function, the high wavenumber
constituents will lag behind, creating the well-known Gibbs phenomenon
which needs to be controlled by a certain amount of viscosity.
Higher order schemes require less viscosity to control the
Gibbs phenomenon \citep{BD02}.
On the other hand, any turbulence simulation requires a sufficient amount
of viscosity to dissipate kinetic energy.
It is therefore thought that for a sixth orders scheme the two limits on
the viscosity are similar and that it is not advantageous to use higher
order representations of the spatial derivatives.

\begin{table}[b!]\caption{
Coefficients $c_j^{(n)}\equiv a_j^{(n)}/b^{(n)}$
}\vspace{12pt}\centerline{\begin{tabular}{rcccccccc}
$N$ & $n$ & $b^{(n)}$ &
$a_0^{(n)}$ &
$a_1^{(n)}$ &
$a_2^{(n)}$ &
$a_3^{(n)}$ &
$a_4^{(n)}$ &
$a_5^{(n)}$ \\
\hline
10 & 1 &  2520 & 0 & 2100 & $-600$ & 150 & $-25$ & 2 \\
 8 & 1 &   840 & 0 &  672 & $-168$ &  32 &  $-3$ &   \\
 6 & 1 &    60 & 0 &   45 &   $-9$ &   1 &       &   \\
 4 & 1 &    12 & 0 &    8 &   $-1$ &     &       &   \\
 2 & 1 &     2 & 0 &    1 &        &     &       &   \\
10 & 2 & 25200 & $-73766$ & 42000 & $-6000$ & $1000$ & $-125$ & $8$ \\
 8 & 2 &  5040 & $-14350$ &  8064 & $-1008$ &  $128$ &   $-9$ &     \\
 6 & 2 &   180 &   $-490$ &   270 &   $-27$ &    $2$ &        &     \\
 4 & 2 &    12 &    $-30$ &    16 &    $-1$ &        &        &     \\
 2 & 2 &     1 &     $-2$ &     1 &         &        &        &     \\
\label{Tcoeff}\end{tabular}}\end{table}

\begin{figure}[t!]\begin{center}
\includegraphics[width=.9\columnwidth]{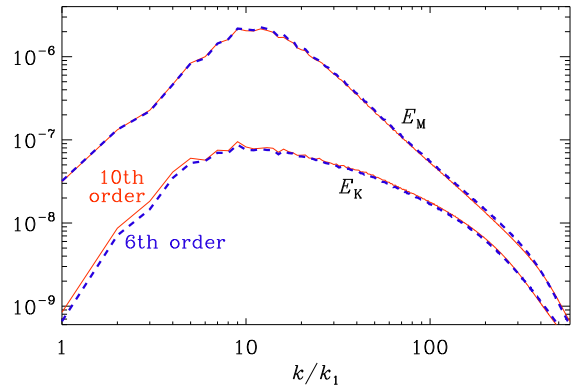}
\end{center}\caption[]{
Magnetic (upper curves) and kinetic (lower curves) energy spectra for
at $t=110$ for the sixth order (blue, dashed) and tenth order (red, solid)
finite difference schemes.
}\label{pspec_comp_10th}\end{figure}

To verify this in the present context, we have run a high Reynolds number case
both with sixth and tenth order schemes.
In the {\sc Pencil Code}, the order of the scheme can easily be changed by
setting {\tt DERIV=deriv\_10th}.
In that case, first and second derivatives are represented as
\EQ
\dd^n f_i/\dd x^n=\sum_{j=-N}^N (\sgn j)^n c_{|j|}^{(n)} f_{i+j}/\delta x^n,
\EN
with coefficient $c_j^{(n)}$ given in \Tab{Tcoeff} for schemes of order $N$.
The result of the comparison is shown in \Fig{pspec_comp_10th}.
The differences between the two cases are negligible, except that with the
more accurate tenth order scheme the inverse transfer of kinetic energy
to larger scales is now slightly stronger.
This is consistent with our earlier findings that the inverse transfer in
nonhelical MHD becomes more pronounced at larger resolution.

\section{Isothermal versus polytropic equation of state}

\begin{figure}[h!]\begin{center}
\includegraphics[width=.9\columnwidth]{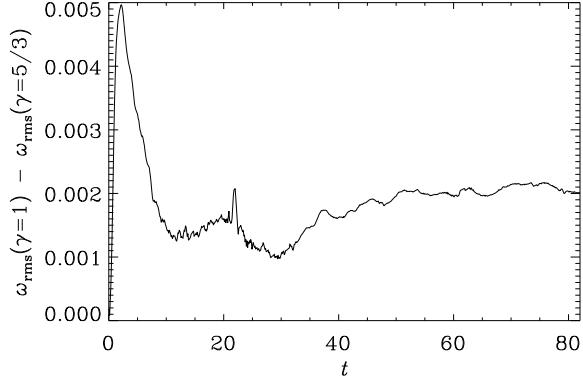}
\end{center}\caption[]{
Difference in rms vorticity, $\orms$, between the isothermal and
polytropic solutions.
}\label{pcomp_orms}\end{figure}

An isothermal equation of state is often used in subsonic compressible
turbulence to approximate the conditions of nearly incompressible flows.
Using instead a polytropic equation of state means that in the momentum
equation the pressure gradient term for an isothermal gas is amended
by a factor $\propto\rho^{\gamma-1}$, i.e.,
\EQ
\cs^2\nab\ln\rho\to\csz^2\left({\rho\over\rho_0}\right)^{\gamma-1}\nab\ln\rho,
\EN
where $\gamma=5/3$ is the polytropic index for a monatomic gas instead
of $\gamma\to1$ for an isothermal gas.
Using $\gamma=5/3$ implies a slightly stiffer equation of state,
so one has to drive stronger to achieve the same compression;
see Sect.~9.3.6 of \cite{Bra03}.
In the present context of subsonic decaying turbulence, this leads to
slightly smaller vorticity fluctuations, as is shown in \Fig{pcomp_orms}.
It is seen that the difference between $\gamma=5/3$ and 1 is negligible
for all practical purposes.

\begin{figure}[t!]\begin{center}
\includegraphics[width=\columnwidth]{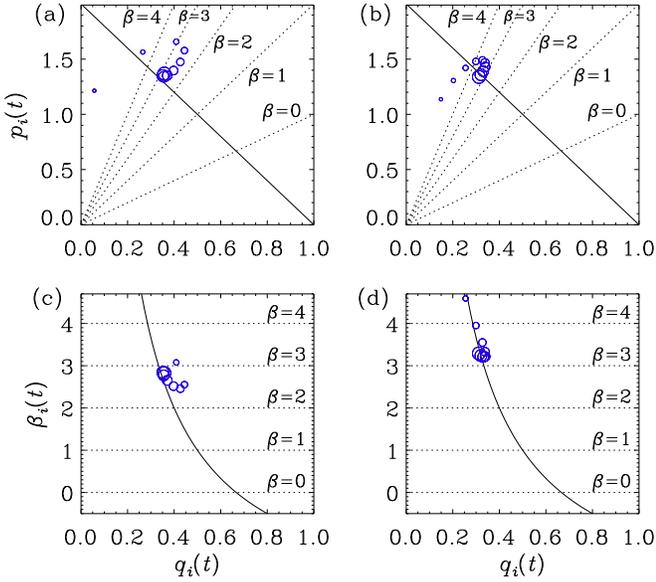}
\end{center}\caption[]{
$pq$ diagrams for hydrodynamic turbulence with $\nu=\const$ (a)
and time-dependent $\nu(t)\propto t^r$ (b) with $r=-0.43$ and $\alpha=4$.
Panels (c) and (d) show the corresponding $\beta q$ diagrams.
Open (closed) symbols corresponds to $i={\rm K}$ (${\rm M}$) and their
sizes increase with time.
}\label{pqhydro_supp}\end{figure}

\begin{figure}[t!]\begin{center}
\includegraphics[width=\columnwidth]{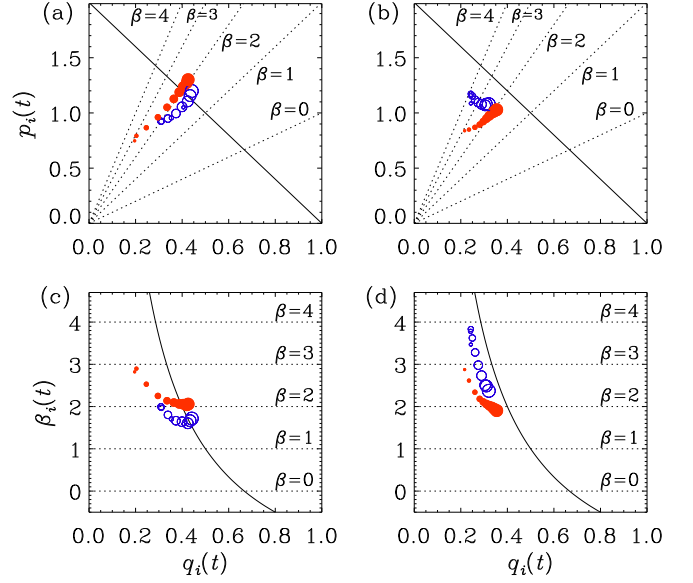}
\end{center}\caption[]{
Similar to \Fig{pqhydro_supp}, but for nonhelical MHD turbulence with
$\nu=\eta=\const$ (a) and time-dependent $\nu(t)=\eta(t)\propto t^r$ (b)
with $r=-0.43$ and $\alpha=4$.
}\label{pqMHD_supp}\end{figure}

\section{Time-dependent $\nu(t)$ and $\eta(t)$}

\begin{table}[b!]\caption{
Exponents $r$ for different $\alpha$.
}\vspace{12pt}\centerline{\begin{tabular}{rcccccccc}
$\alpha$ &   0  & $1\quad$ &     2    &    3    &     4   \\
\hline
$r$      &$\quad0.33\quad$&$0\quad$& $-0.20\quad $ & $-0.33\quad$ & $-0.43\quad$ \\
\label{TRofAlp}\end{tabular}}\end{table}

\begin{figure}[t!]\begin{center}
\includegraphics[width=\columnwidth]{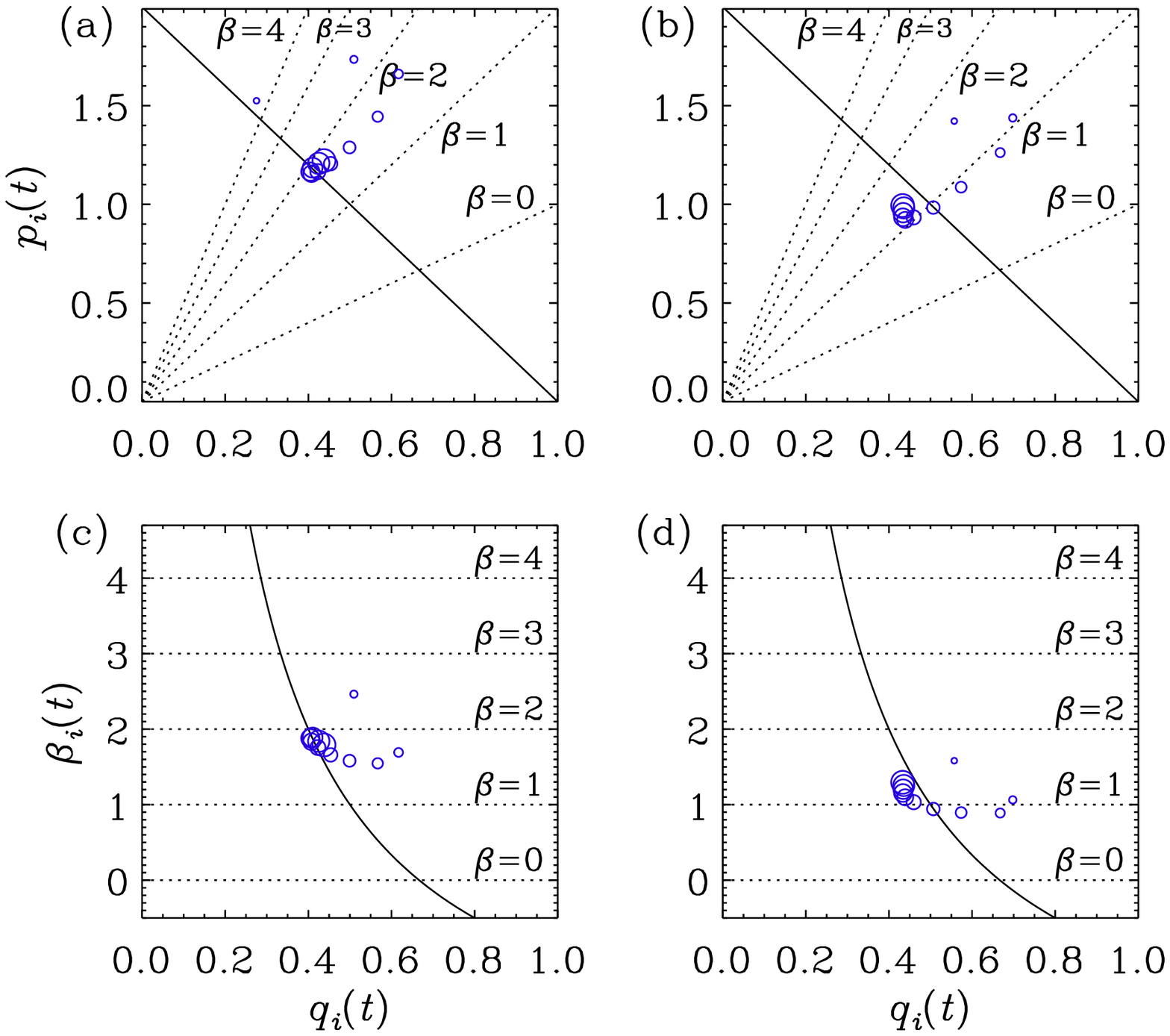}
\end{center}\caption[]{
Similar to \Fig{pqhydro_supp}, but for $\alpha=2$ (a) and $\alpha=1$ (b)
with $\nu=\const$.
}\label{pqhydro_alp_supp}\end{figure}

As pointed out by Olesen \cite{Ole97}, the hydrodynamic and MHD
equations are invariant under rescaling $x\to\tilde{x}\ell$ and
$t\to\tilde{t}\ell^{1/q}$ provided also $\nu$ and $\eta$ are
being dynamically rescaled such that
\EQ
\nu(t)=\nu_0\,\max(t/t_0,1)^r,\quad
\eta(t)=\eta_0\,\max(t/t_0,1)^r,
\label{UseMax}
\EN
with
\EQ
r=2q-1=(1-\alpha)/(3+\alpha);
\EN
see \Tab{TRofAlp}.
The use of the $\max$ function in \Eq{UseMax} limits the values of
$\nu\leq\nu_0$ and $\eta\leq\eta_0$ for $t\leq t_0$ when $r<0$.
At large Reynolds numbers, the time-dependence is not expected to be
important.
To verify this, we compare in \Fig{pqhydro_supp} hydrodynamic runs
with constant and time-dependent $\nu$ using $\alpha=4$.
Both cases are similar and the case with time-dependent $\nu$
still has $\beta=3\neq\alpha$.
Similar behavior is found in MHD; see \Fig{pqMHD_supp},
where we compare runs with constant and time-dependent $\nu$ and $\eta$
using again $\alpha=4$.
In both cases, we find $\beta=2\neq\alpha$.

In agreement with earlier work we find that in hydrodynamic cases
with $\alpha=2$ and $\alpha=1$, we have $\beta=\alpha$ \cite{YHB04}.
This is demonstrated in \Fig{pqhydro_alp_supp}, where we show the $pq$
and $\beta q$ diagrams for these two cases.

\end{document}